\documentclass[%
 aip,
 amsmath,amssymb,
reprint,%
floatfix,
]{revtex4-1}

\usepackage{graphicx}
\usepackage{dcolumn}
\usepackage{comment}
\usepackage{bm}
\usepackage{multirow}
\usepackage{csquotes}
\usepackage{algpseudocode}
\usepackage[mathscr]{euscript}
\usepackage{amsmath}
\usepackage{bbm}
\usepackage{xcolor}

\newcommand\bra[1] {\langle {#1} |}
\newcommand\ket[1] {| {#1} \rangle}
\newcommand\braket[2] {\langle {#1} | {#2} \rangle}

\newcommand{\themi}{M_I}
\newcommand{\thezi}{Z_I}

\newcommand{\theri}{\mathbf{R}_I}

\newcommand{\theb}{\mathbf{B}}

\newcommand{\thennuc}{N_\mathrm{nuc}}

\newcommand{\theoph}{H}

\def\sa{\tau}
\def\sb{\kappa}
\def\sc{\xi}
\def\sd{\eta}
\def\eaoa{\mu}
\def\eaob{\nu}
\def\eaoc{\lambda}
\def\eaod{\gamma}
\def\eaosa{\eaoa^{\sa}}
\def\eaosb{\eaob^{\sb}}
\def\eaosc{\eaoc^{\sc}}
\def\eaosd{\eaod^{\sd}}
\def\paoa{\alpha}
\def\paob{\beta}
\def\paoc{\chi}
\def\paod{\delta}
\def\paosa{\paoa^{\sa}}
\def\paosb{\paob^{\sb}}
\def\paosc{\paoc^{\sc}}
\def\paosd{\paod^{\sd}}
\def\genexc{\tau}
\def\genoa{\bar{P}}
\def\genob{\bar{Q}}
\def\genvec{\mathbf{r}}
\def\exclaba{\tilde{\nu}_1}
\def\exclabb{\tilde{\nu}_2}
\def\exclabc{\tilde{\nu}_3}
\newcommand\rpamat[3]{\mathbf{#1}^{#2}_{#3}}
\def\elab{\text{e}}
\def\eelab{\text{ee}}
\def\plab{\text{p}}
\def\pplab{\text{pp}}
\def\eplab{\text{ep}}
\def\pelab{\text{pe}}
\def\ilab{\text{i}}
\newcommand{\theG}{\mathbf{G}}
\newcommand{\thenel}{N_\mathrm{el}}
\newcommand{\thenp}{N_{\plab}}
\newcommand{\theari}{\mathbf{A}(\mathbf{r}^{\elab}_i)}
\newcommand{\thearip}{\mathbf{A}(\mathbf{r}^{\plab}_{I})}
\newcommand{\thepiel}{\mathbf{p}^{\elab}_i}
\newcommand{\thepip}{\mathbf{p}^{\plab}_{I}}

\begin{document}

\title{Time-Dependent Nuclear-Electronic Orbital Hartree--Fock Theory in a Strong Uniform Magnetic Field
}

\author{Tanner Culpitt}
\email{t.p.culpitt@kjemi.uio.no}
\affiliation
{Hylleraas Centre for Quantum Molecular Sciences,  Department of Chemistry, 
University of Oslo, P.O. Box 1033 Blindern, N-0315 Oslo, Norway}
\author{Laurens D. M. Peters}
\affiliation
{Hylleraas Centre for Quantum Molecular Sciences,  Department of Chemistry, 
University of Oslo, P.O. Box 1033 Blindern, N-0315 Oslo, Norway}
\author{Erik I. Tellgren}
\affiliation
{Hylleraas Centre for Quantum Molecular Sciences,  Department of Chemistry, 
University of Oslo, P.O. Box 1033 Blindern, N-0315 Oslo, Norway}
\author{Trygve Helgaker}
\affiliation
{Hylleraas Centre for Quantum Molecular Sciences,  Department of Chemistry, 
University of Oslo, P.O. Box 1033 Blindern, N-0315 Oslo, Norway}

\begin{abstract}
The Born--Oppenheimer (BO) approximation is less accurate in the presence of a strong magnetic field than in the absence of a field. This is due to the complicated and unpredictable response of electronic structure to the field, especially in the mixed regime $B \approx B_0 = 2.35 \times 10^5\,$T. Therefore, it is desirable to explore non-BO methods in magnetic fields. In this work, the nuclear--electronic orbital (NEO) method is employed to study protonic vibrational excitation energies in the presence of a strong magnetic field. NEO Generalized Hartree--Fock theory and time-dependent Hartree--Fock theory are derived and implemented, accounting for all terms that result as a consequence of the nonperturbative treatment of molecular systems in a magnetic field. The NEO results for HCN and FHF$^-$ with clamped heavy nuclei are compared against the quadratic eigenvalue problem (QEP). Each molecule has three semi-classical modes owing to the hydrogen -- two precession modes that are degenerate in the absence of a field and one stretching mode. The NEO-TDHF model is found to perform well -- in particular it automatically captures the screening effects of the electrons on the nuclei, which are quantified through the difference in energy of the precession modes. 
 \end{abstract}

\maketitle

\section{Introduction}

There has been recent interest in molecular electronic structure\cite{Lange2012,Tellgren2012,Stopkowicz2015,FURNESS_JCTC11_4169, Hampe2017, Stopkowicz2018, Sun2019, Hampe2019, Sen2019, Hampe2020, Irons2021,Wibowo2021,David2021} and dynamics\cite{Ceresoli2007,Culpitt2021,Peters2021,Monzel2022} for molecules in a strong magnetic field. Introduction of a magnetic field presents new challenges for the derivation and implementation of quantum-chemical methods. A nonperturbative magnetic treatment of electronic structure requires accounting for additional terms in the electronic Hamiltonian that give rise to orbital- and spin-Zeeman effects. Additionally, all mathematics and implementation are necessarily complex. For Born--Oppenheimer (BO) molecular dynamics, new terms arise in the nuclear equations of motion due to the magnetic vector potential and geometric vector potential, with the latter potential being related to a geometric or Berry phase.\cite{Berry1984,Mead1992,Anandan1997,Resta2000} In the context of dynamics, these potentials give rise to the Lorentz force acting on the nuclei as well as a Berry (screening) force,\cite{Ceresoli2007,Culpitt2021,Peters2021,Monzel2022,Bian_JCP2022} which is given in terms of the Berry curvature.\cite{Resta2000,Ceresoli2007,Culpitt2021,Peters2021}

Most methods for studying nonperturbative electronic structure and dynamics in a magnetic field have been developed within the BO paradigm. However, in a strong magnetic field, the ordering of electronic energy levels can change substantially.\cite{Wunner_1987} It is therefore plausible that the BO approximation is generally less legitimate in a strong magnetic field than in the field-free case. For this reason, it is desirable to investigate nonperturbative, non-BO electronic structure methods for calculating molecular properties in the presence of a magnetic field. The non-BO theory of molecular systems in a magnetic fields has been studied,\cite{Avron1978,Schmelcher1988,Schmelcher_1988-2,Detmer_1995, Schmelcher1997} but general application to molecular systems is not widespread. Recent progress has been made by Adamowicz \emph{et al.}\cite{Adamowicz2015,Adamowicz2017,Adamowicz2018,Adamowicz2020,Adamowicz2022} studying the HD molecule beyond the BO approximation in a magnetic field using explicitly correlated Gaussians.

There are several methods for calculating non-BO molecular properties that fall under the umbrella of multicomponent quantum theory,\cite{Kreibich2001,Bochevarov2004,Nakai2007,Ishimoto2009,Reyes2019,Pavosevic2020} where ``multicomponent" refers to treating more than one type of particle quantum mechanically, all on equal footing. In practice, the central theme of many of these methods is treating both electrons and nuclei quantum mechanically with orbital-based techniques, an idea dating back at least to the work of Thomas\cite{Thomas_1969} in 1969 and Thomas and Joy\cite{Thomas_1970} in 1970. One such method is the nuclear--electronic orbital (NEO) method, which has historically been used to treat all electrons and selected nuclei (typically protons) quantum mechanically in a molecular system. This approach eliminates difficulties associated with molecular rotations and translations since the molecular frame is fixed by the clamped nuclei. The NEO method may thus be described as ``partially" non-BO, representing a compromise between a fully quantum treatment and a partial quantum treatment that can be used to study chemically relevant or interesting quantum effects of certain nuclei. 

Research related to the NEO method is ongoing, and in the previous decades there have been many advances. For wave-function theory, these include NEO Hartree--Fock (NEO-HF) theory,\cite{Webb2002} NEO time-dependent Hartree--Fock (NEO-TDHF) theory,\cite{Yang2018-1} NEO full-configuration-interaction (NEO-FCI) theory,\cite{Webb2002} NEO explicitly-correlated Hartree--Fock (NEO-XCHF) theory,\cite{Swalina2006, Sirjoosingh2013} NEO multiconfigurational self-consistent field  (NEO-MCSCF) theory,\cite{Webb2002} and NEO coupled-cluster (NEO-CC) theory.\cite{Pavosevic2019,Pavosevic2019-2} Additionally, there have also been advances in NEO density-functional theory (NEO-DFT),\cite{Yang2017,Brorsen2017,Tao2019} time-dependent density-functional theory (NEO-TDDFT),\cite{Yang2018-1} NEO real-time TDDFT,\cite{Zhao2020_RTneo} as well as molecular dynamics within the NEO framework.\cite{Zhao2020,Zhao2021,Tao2021} Finally, a new variant of the NEO method, denoted constrained NEO (cNEO),\cite{Xu2020-1,Xu2020-2,Xu_JACS2022} has recently been developed, allowing for the fully quantum treatment of all nuclei by constraining the expectation values of nuclear densities.

Here, we concern ourselves with the development and application of nonperturbative variants of NEO-HF and NEO-TDHF theories that will be relevant to molecular systems in a strong magnetic field, both methods having been programmed in the software package LONDON.\cite{LondonProgram} In particular, we are interested in NEO-TDHF vibrational excitation energies and how they are affected by the introduction of a magnetic field. This has direct relevance to the results obtained using BO molecular dynamics for molecules, including the effects of the Lorentz force and Berry curvature on nuclei. In principle, these effects should be captured directly with the NEO model. To what extent the nuclear Lorenz force will be accurately screened by the electrons within the NEO paradigm is an interesting and open question that we seek to investigate. Toward this end, we examine the HCN and FHF$^-$ molecules, which are both well known systems in the NEO context, having been studied extensively.\cite{Yang2018-1,Yang2018-2,Culpitt2019,Pavosevic2020} 

This work is organized as follows. Section II contains theoretical background and derivation pertaining to the NEO Generalized Hartree--Fock (NEO-GHF) method, as well as a derivation of the generally complex NEO-TDHF working equations. A comparison is made with the results from the quadratic eigenvalue problem in a magnetic field. Section III presents protonic vibrational excitation energies for the HCN and FHF$^-$ molecules as a function of magnetic field strength. The work is summarized and future directions are given in Section~IV.

\section{Theory}

We consider a joint system of $\thennuc$ classical/clamped nuclei, $\thenp$ quantum protons, and $\thenel$ electrons. We use the notation $\themi$, $\thezi$, and $\theri$ for the  mass, atom number, and position of clamped nucleus $I$, respectively. We use $\mathbf{r}^{\elab}_i$ and $\mathbf{p}^{\elab}_i$ for the position operator and canonical momentum operator of electron $i$, respectively. We use $\mathbf{r}^{\plab}_{I}$ and $\mathbf{p}^{\plab}_{I}$ for the position  and canonical-momentum operators of proton $I$, respectively.
The vectors of collective clamped nuclear, electronic, and protonic coordinates are denoted by $\mathbf{R}$, $\mathbf {r}^{\elab}$, and $\mathbf {r}^{\plab}$, respectively. The vector potential of a uniform magnetic field $\mathbf B$ at position $\mathbf u$ is given by $\mathbf A(\mathbf u) = \frac{1}{2} \theb \times (\mathbf u-\theG)$, where $\theG$ is the gauge origin. 

\subsection{The NEO Hamiltonian in a uniform magnetic field}

The nonrelativistic Schr\"odinger--Pauli Hamiltonian of a molecular system within the NEO framework in a uniform magnetic field comprised of heavy nuclei with quantum protons and electrons can be written as
\begin{align}
    \theoph_\text{NEO} = H^{\elab} + H^{\plab} + V^{\eplab} + V^{\text{nuc}}, \label{ham_neo}
\end{align}
where we have introduced the electronic Hamiltonian with terms representing the electronic kinetic energy, the repulsion between the electrons, and the attraction of the electrons to the clamped nuclei,
\begin{align}
H^{\elab} &= \frac{1}{2m^{\elab}} \sum_{i=1}^{\thenel} [\bm{\sigma}\cdot(\thepiel - q^{\elab} \theari)]^2 \nonumber \\ 
 & \quad + 
  \sum_{i>j=1}^{N_\text{el}}\frac{e^2}{4 \pi \varepsilon_0\vert \mathbf r^{\elab}_i - \mathbf r^{\elab}_j\vert} - \sum_{i=1}^{N_\text{el}}\sum_{I=1}^{N_\text{nuc}}\frac{Z_I e^2}{4 \pi \varepsilon_0\vert \mathbf r^{\elab}_i - \mathbf R_I\vert}, \label{ham_e}
  \end{align}
  the protonic Hamiltonian with similar terms for the protons,
  \begin{align}
  H^{\plab} &= \frac{1}{2m^{\plab}} \sum_{I=1}^{\thenp} [\bm{\sigma}\cdot(\thepip - q^{\plab} \thearip)]^2 \nonumber \\ 
 & \quad + 
  \sum_{I>J=1}^{N_\text{p}}\frac{Z_{I}Z_{I}e^2}{4 \pi \varepsilon_0\vert \mathbf r^{\plab}_{I} - \mathbf r^{\plab}_{J}\vert} + \sum_{J=1}^{N_\text{p}}\sum_{I=1}^{N_\text{nuc}}\frac{Z_{J}Z_I e^2}{4 \pi \varepsilon_0\vert \mathbf r^{\plab}_{J} - \mathbf R_I\vert}, \label{ham_p}
\end{align}
the electron--proton attraction operator,
\begin{align}
 V^{\eplab} &= - \sum_{J=1}^{N_\text{p}}\sum_{i=1}^{N_\text{el}}\frac{Z_{J} e^2}{4 \pi \varepsilon_0\vert \mathbf r^{\plab}_{J} - \mathbf r^{\elab}_i\vert}, \label{Vep}
 \end{align}
 and the operator representing the
 repulsion between the clamped nuclei,
 \begin{align}
 V^{\text{nuc}} &= \sum_{I>J=1}^{N_\text{nuc}}\frac{Z_I Z_J e^2}{4 \pi \varepsilon_0\vert \mathbf R_I - \mathbf R_J\vert}. \label{vnuc}
\end{align}
In Eqs.~\eqref{ham_e}--\eqref{vnuc}, $m^{\elab}$ is the electron mass, $m^{\plab}$ is the proton mass, $e$ is the elementary charge, $\varepsilon_0$ is the vacuum permittivity, $q^{\elab} = -e$ is the electron charge, $q^{\plab} = e$ is the proton charge, and $\bm{\sigma}$ is the vector of Pauli matrices
\begin{align}
\bm{\sigma}_x = \begin{pmatrix}
0 & 1 \\
1 & 0
\end{pmatrix}, \;
\bm{\sigma}_y = \begin{pmatrix}
0 & -\ilab \\
\ilab & 0
\end{pmatrix}, \;
\bm{\sigma}_z = \begin{pmatrix}
1 & 0 \\
0 & -1
\end{pmatrix}.
\end{align}
In the NEO Born--Oppenheimer approximation, the total ground-state wave function associated with the electrons and quantum protons can be written as the product
\begin{align}
\Psi(\mathbf{r}^{\elab},\mathbf{r}^{\plab},\mathbf{R}) = \psi^{\elab}(\mathbf{r}^{\elab};\mathbf{R})\psi^{\plab}(\mathbf{r}^{\plab};\mathbf{R})\Theta(\mathbf{R})
\label{wf_def} \ ,
\end{align}
where $\psi^{\elab}(\mathbf{r}^{\elab};\mathbf{R})$ is the electronic wave function, $\psi^{\plab}(\mathbf{r}^{\plab};\mathbf{R})$ is the protonic wave function, and $\Theta(\mathbf{R})$ is the wave function for the heavy nuclei. From here onward, we suppress the arguments of the wave functions. Note that the forms of the electronic and protonic kinetic energy operators will require additional terms to enter into the one-particle operators of Hartree--Fock theory, and also necessitate a generally complex implementation in addition to compensation for the gauge-dependence of the Hamiltonian.

\subsection{Quadratic eigenvalue problem}

For conventional electronic-structure theory within the BO approximation, the recently developed theory of molecular vibrations in the presence of a magnetic field~\cite{Tellgren_velvib_arxiv2022} provides a baseline for a comparison with the NEO-TDHF results. The magnetic field introduces velocity-dependent forces and the classical equations of motion for the nuclei become
\begin{equation}
    m_I \ddot{\mathbf{R}}_I = - \boldsymbol \nabla_I V(\mathbf R) 
    + q_I \dot{\mathbf{R}}_I\times\mathbf{B} + \sum_J \boldsymbol{\Omega}^{\text{int}}_{IJ}(\mathbf{R}) \, \dot{\mathbf{R}}_J. \label{newton_eqn}
\end{equation}
Here, the first term $-\boldsymbol \nabla_I V(\mathbf{R})$ is the BO potential-energy force on nucleus $I$, the second term $q_I \dot{\mathbf{R}}_I\times\mathbf{B}$ is the (bare) Lorentz force on nucleus $I$, while the last term $\sum_J \boldsymbol{\Omega}^{\text{int}}_{IJ}(\mathbf{R}) \dot {\mathbf R}_J$ is the Berry force on nucleus $I$, expressed in terms of the Berry curvature $\boldsymbol{\Omega}^{\text{int}}_{IJ}(\mathbf{R})$, whose elements are given by\cite{Culpitt2021,Culpitt2022}
\begin{align}
    \Omega_{I\alpha J\beta}^{\text{int}} &= \ilab \hbar
    \big[
    \braket{\nabla_{I\alpha} \psi^{\elab}}{\nabla_{J\beta} \psi^{\elab}} -  \braket{\nabla_{J\beta} \psi^{\elab}}{\nabla_{I\alpha} \psi^{\elab}}
    \big] \nonumber \\
&= -2\hbar\text{Im}\braket{\nabla_{I\alpha} \psi^{\elab}}{\nabla_{J\beta} \psi^{\elab}},
    \label{berry_curv}
\end{align}
where $I\alpha$ is a composite nuclear--Cartesian index and $\nabla_{I\alpha}$ differentiates with respect to $R_{I\alpha}$. The Berry force is also referred to as the screening force, because it represents the screening (due to the electrons) of the bare Lorentz force acting on the nuclei. Introducing the `external Berry curvature'
\begin{align}
\Omega^{\text{ext}}_{I\alpha,J\beta} = \delta_{IJ} q_I \epsilon_{\alpha\beta\zeta} B_{\zeta},
\label{omega_lorentz}
\end{align}
we may write the the nuclear equations of motion more compactly as
\begin{equation}
    m_I \ddot{\mathbf{R}}_I = - \boldsymbol \nabla_I V(\mathbf R) 
    + \sum_J \boldsymbol{\Omega}^{\text{tot}}_{IJ}(\mathbf{R}) \, \dot{\mathbf{R}}_J,
\end{equation}
in terms of the total Berry curvature tensor
\begin{align}
\boldsymbol{\Omega}^{\text{tot}}_{IJ}(\mathbf{R}) = \boldsymbol{\Omega}^{\text{ext}}_{IJ}(\mathbf{R}) + \boldsymbol{\Omega}^{\text{int}}_{IJ}(\mathbf{R}). 
\label{omega_tot}
\end{align}

Truncating the equation of motion to first order in the displacement $\boldsymbol{\eta} = \mathbf{R} - \mathbf{R}^{\text{eq}}$ from a minimum on the potential energy surface (PES) $V(\mathbf{R}^{\text{eq}})$ and transforming
to the frequency domain under the convention that
\begin{align}
    \label{eqFourierConvention}
    F(\omega) = \int_{-\infty}^{\infty} \!\!f(t) \mathrm e^{\ilab\omega t} \,\mathrm dt,  \\
    f(t)=\frac{1}{2\pi}\int_{-\infty}^{\infty}\!\!F(\omega) \mathrm e^{-\ilab\omega t} \,\mathrm d\omega ,
\end{align}
we obtain
\begin{align}
    \label{eqSemiClassQEP}
    \omega^2 \, M_I \boldsymbol{\eta}_I (\omega) &= \sum_J \mathbf{H}_{IJ}(\mathbf{R}^{\text{eq}}) \,\boldsymbol{\eta}_J(\omega) \nonumber \\ & \quad + \ilab\omega \sum_J \boldsymbol{\Omega}^{\text{tot}}_{IJ}(\mathbf{R}^{\text{eq}}) \, \boldsymbol{\eta}_J(\omega).
\end{align}
where the elements of the Hessian matrix at $\mathbf{R}^{\text{eq}}$ are given by
\begin{align}
H_{I\alpha,J\beta}(\mathbf{R}^{\text{eq}}) &= \left.\nabla_{I\alpha}\nabla_{J\beta} V(\mathbf{R})\right\vert_{\mathbf{R}=\mathbf{R}^{\text{eq}}}.
\end{align}
Neglecting the contribution from the total Berry curvature, we recover the usual eigenvalue problem for molecular vibrations, whose eigenvalue is the squared oscillation frequency $\omega^2$. With Berry curvature included, we have a \emph{quadratic eigenvalue problem (QEP)},~\cite{Tellgren_velvib_arxiv2022,TISSEUR_SIAMR43_235} which yields the oscillation frequency $\omega$ directly rather than $\omega^2$. Note that the external and internal contributions to the total Berry curvature in Eq.~\eqref{omega_tot} may be turned on and off separately, thereby allowing for the investigation of different cases where the Lorentz force or Berry force or both are set to zero in the QEP. For a complete exposition on the QEP see Ref.~\onlinecite{Tellgren_velvib_arxiv2022}. For more information on the Berry curvature and molecular dynamics in magnetic fields, see Refs.~\onlinecite{Peters2021,Culpitt2021,Culpitt2022,Monzel2022}.

\subsection{NEO-GHF theory}

In this section, we present the equations for generalized NEO Hartree--Fock (NEO-GHF) theory in a uniform magnetic field. This model subsumes all other spin variants of Hartree--Fock theory, such as unrestricted and restricted Hartree--Fock theories, as the latter two may be derived from the GHF equations by imposing appropriate restrictions on spin. 

In what follows, lower (upper) case $p,q,r,s$ indices refer to general electronic (protonic) spinors, lower (upper) case $i,j,k,l$ indices refer to occupied electronic (protonic) spinors, and lower (upper) case $a,b,c,d$ indices refer to virtual electronic (protonic) spinors. Throughout this section, Greek subscript indices $\eaoa,\eaob,\eaoc,\eaod$ refer to the electronic atomic-orbital (AO) basis, while Greek subscript indices $\paoa$,$\paob$,$\paoc$,$\paod$ refer to the protonic AO basis. Superscript indices $\sa$,$\sb$,$\sc$,$\sd$ all refer to spin, which is to say $\sa,\sb,\sc, \sd \ \in \ \{\uparrow,\downarrow\}$.

A generic spinor is a linear combination of spin-dependent basis functions $\phi(\mathbf{x})$ according to
\begin{align}
\Phi_{i}(\mathbf{x}) 
= \sum _{\mu\tau} c^{\tau}_{\mu i}\phi_{\mu}^{\tau}(\mathbf{x})
\label{e_spinor} \ ,
\end{align}
where $\mathbf{x}$ is a mixed space--spin coordinate, and each $\phi_{\mu}^{\tau}(\mathbf{x})$ is the product of a spatial function $\theta_{\mu}(\mathbf{r})$ and a spin function $\tau$. Thus, for $N$ basis functions,\textbf{} there are 2$N$ terms in the sum in Eq.\,\eqref{e_spinor} and the Fock and density matrices are blocked 2$N\times$2$N$ matrices in the AO basis.
We adopt the chemists' notation for two-particle integrals over spin-dependent basis functions
\begin{align}
&(\phi_\mu^{\tau}\phi_\nu^{\kappa}\vert\phi_\lambda^{\xi}\phi_\gamma^{\eta}) = (\eaosa\eaosb \vert \eaosc\eaosd) \nonumber \\
&=\int \!\mathrm  d\mathbf{x}_1 \mathrm d\mathbf{x}_2\phi_{\mu}^{\tau *}(\mathbf{x}_1)\phi_{\nu}^{\kappa}(\mathbf{x}_1)r_{12}^{-1}\phi_{\lambda}^{\xi *}(\mathbf{x}_2)\phi_{\gamma}^{\eta}(\mathbf{x}_2) \nonumber \\
&=\int \!\mathrm  d\mathbf{r}_1 \mathrm d\mathbf{r}_2\theta_{\mu}^{*}(\mathbf{r}_1)\theta_{\nu}(\mathbf{r}_1)r_{12}^{-1}\theta_{\lambda}^{*}(\mathbf{r}_2)\theta_{\gamma}(\mathbf{r}_2)
\label{cphf_tpi}\delta_{\tau\kappa}\delta_{\xi\eta},
\end{align}
and adopt the Coulomb--exchange shorthand notation
\begin{align}
(\eaosa\eaosb\Vert \eaosc\eaosd)= (\eaosa\eaosb\vert \eaosc\eaosd) -
(\eaosa\eaosd\vert \eaosc\eaosb)
\label{cphf_tpi1} \ .
\end{align}
Taking the electronic and protonic wave functions in Eq.\,\eqref{wf_def} to be Slater determinants comprised of spinors, the NEO-GHF energy in a uniform magnetic field becomes
\begin{align}
E &= E^{\elab} + E^{\plab} + E^{\eplab} + V^{\text{nuc}}, \label{hf_energy}
\end{align}
where the electronic, protonic and electronic--protonic energy contributions, respectively, in the AO basis are given by
\begin{align}
E^{\elab} &= \sum_{\sa\sb}\sum_{\eaoa\eaob}D^{\elab,\sb\sa}_{\eaob\eaoa}h_{\eaoa\eaob}^{{e,\sa\sb}} \nonumber \\
& \quad + \frac{1}{2}\sum_{\sa\sb\sc\sd}\sum_{\eaoa\eaob\eaoc\eaod}D^{\elab,\sb\sa}_{\eaob\eaoa}D^{\elab,\sd\sc}_{\eaod\eaoc}\,(\eaosa\eaosb\Vert \eaosc\eaosd) \nonumber \\
& \quad +  \frac{1}{2m^{\elab}}\sum_{\Upsilon}\sum_{\eaoa\eaob}B_{\Upsilon} \mathscr{S}^{\elab}_{\eaoa\eaob}\text{Tr}_{\Sigma}[\bm{\sigma}_{\Upsilon}\,\mathbf{D}^{\elab}]_{\eaob\eaoa}, 
\label{hf_energy_ee} 
\end{align}
\begin{align}
E^{\plab} &= \sum_{\sa\sb}\sum_{\paoa\paob}D^{\plab,\sb\sa}_{\paob\paoa}h_{\paoa\paob}^{\plab,\sa\sb} \nonumber \\
& \quad + \frac{1}{2}\sum_{\sa\sb\sc\sd}\sum_{\paoa\paob\paoc\paod}D^{\plab,\sb\sa}_{\paob\paoa}D^{\plab,\sd\sc}_{\paod\paoc}(\paosa\paosb\Vert \paosc\paosd) \nonumber \\
& \quad - \frac{1}{2m^{\plab}}\sum_{\Upsilon}\sum_{\paoa\paob}B_{\Upsilon} \,\mathscr{S}^{\plab}_{\paoa\paob}\text{Tr}_{\Sigma}[\bm{\sigma}_{\Upsilon}\mathbf{D}^{\plab}]_{\paob\paoa}, \label{hf_energy_pp}
\end{align}
\begin{align}
E^{\eplab} &= - \sum_{\sa\sb\sc\sd}\sum_{\eaoa\eaob}\sum_{\paoa\paob}D^{\elab,\sb\sa}_{\eaob\eaoa}D^{\plab,\sd\sc}_{\paob\paoa}\,(\eaosa\eaosb\vert \paoa^{\sc}\paob^{\sd}).
\label{hf_energy_ep}
\end{align}
In these expressions, $\text{Tr}_{\Sigma}$ denotes the trace over the spin degrees of freedom,  $h_{\eaoa\eaob}^{\elab,\sa\sb}$ and $h_{\paoa\paob}^{\plab,\sa\sb}$ are, respectively, the one-electron and one-proton Hamiltonian matrix elements in the presence of a magnetic field,  $B_\Upsilon$ is a Cartesian component of the magnetic field vector, $\mathscr{S}^{\elab}$ and $\mathscr{S}^{\plab}$  are the \textit{purely spatial} electronic and protonic overlap matrices, respectively,
\begin{align}
\mathscr{S}^{\elab}_{\eaoa\eaob} &= \int\! \mathrm d\mathbf{r}^{\elab} \theta^{\elab *}_{\eaoa}(\mathbf{r}^{\elab})\theta^{\elab}_{\eaob}(\mathbf{r}^{\elab}), \\
\mathscr{S}^{\plab}_{\paoa\paob} &= \int\! \mathrm d\mathbf{r}^{\plab} \theta^{\plab *}_{\paoa}(\mathbf{r}^{\plab})\theta^{\plab}_{\paob}(\mathbf{r}^{\plab}),
\label{cphf_overlap}
\end{align}
while $\mathbf{D}^{\elab}$ and $\mathbf{D}^{\plab}$ are, 
respectively, the electronic and protonic spin-blocked density matrices
\begin{align}
D^{\elab,\sa\sb}_{\eaoa\eaob} &= \sum_{i}c^{\elab,\sa}_{\eaoa i}c^{\elab,\sb *}_{\eaob i}, \label{P_def_e} \\
D^{\plab,\sa\sb}_{\paoa\paob} &= \sum_{I}c^{\plab,\sa}_{\paoa I}c^{\plab,\sb *}_{\paob I}
\label{P_def_p}.
\end{align}
Note that the one-electron and one-proton Hamiltonian matrix elements contain their respective contributions from the canonical momentum, external potential, orbital-Zeeman and diamagnetic terms.\cite{Sen2018,Sun2019} The orbital-Zeeman and diamagnetic contributions are due to the presence of the magnetic vector potential in the Hamiltonian, with opposite signs arising from the opposite
charges $q^{\elab}$ and $q^{\plab}$ in Eq.~\eqref{ham_neo}. 

Minimization of the energy with respect to the electronic and protonic spinor coefficients under the constraint of orthonormalization of the spinors gives rise to the coupled NEO-GHF Roothaan--Hall equations,
\begin{align}
\mathbf{F}^{\elab}\mathbf{C}^{\elab} &= \mathbf{S}^{\elab}\mathbf{C}^{\elab}\mathbf{\epsilon}^{\elab}
\label{RH_e} \ , \\
\mathbf{F}^{\plab}\mathbf{C}^{\plab} &= \mathbf{S}^{\plab}\mathbf{C}^{\plab}\mathbf{\epsilon}^{\plab},
\label{RH_p}
\end{align}
which may be written in spin-blocked form according to
\begin{align}
\begin{pmatrix}
\mathbf{F}^{\elab,\uparrow\uparrow} & \mathbf{F}^{\elab,\uparrow\downarrow} \\
\mathbf{F}^{\elab,\downarrow\uparrow} & \mathbf{F}^{\elab,\downarrow\downarrow}
\end{pmatrix} 
\begin{pmatrix}
\mathbf{c}^{\elab,\uparrow}  \\
\mathbf{c}^{\elab,\downarrow} 
\end{pmatrix} 
&= \begin{pmatrix}
\mathbf{S}^{\elab,\uparrow\uparrow} & \mathbf{0} \\
\mathbf{0} & \mathbf{S}^{\elab,\downarrow\downarrow}
\end{pmatrix} 
\begin{pmatrix}
\mathbf{c}^{\elab,\uparrow}  \\
\mathbf{c}^{\elab,\downarrow} 
\end{pmatrix} \bm{\epsilon}^{\elab}, \label{RH_sb_e}
\end{align}
\begin{align}
\begin{pmatrix}
\mathbf{F}^{\plab,\uparrow\uparrow} & \mathbf{F}^{\plab,\uparrow\downarrow} \\
\mathbf{F}^{\plab,\downarrow\uparrow} & \mathbf{F}^{\plab,\downarrow\downarrow}
\end{pmatrix} \!
\begin{pmatrix}
\mathbf{c}^{\plab,\uparrow}  \\
\mathbf{c}^{\plab,\downarrow} 
\end{pmatrix} 
\!&=\! \begin{pmatrix}
\mathbf{S}^{\plab, \uparrow\uparrow} & \mathbf{0} \\
\mathbf{0} & \mathbf{S}^{\plab,\downarrow\downarrow}
\end{pmatrix} 
\begin{pmatrix}
\mathbf{c}^{\plab,\uparrow}  \\
\mathbf{c}^{\plab,\downarrow} 
\end{pmatrix}\! \bm{\epsilon}^{\plab}.
\label{RH_sb_p}
\end{align}
In these expressiones, the spin-blocked overlap matrices are given by
\begin{align}
S^{\elab,\sa\sb}_{\eaoa\eaob} &= \int\! \mathrm d\mathbf{x}\, \phi^{\elab,\sa *}_{\eaoa}(\mathbf{x})\,\phi^{\elab,\sb}_{\eaob}(\mathbf{x})
\label{S_e}, \\
S^{\plab,\sa\sb}_{\paoa\paob} &= \int\! \mathrm d\mathbf{x} \phi^{\plab,\sa *}_{\paoa} (\mathbf{x})\phi^{\plab,\sb}_{\paob}(\mathbf{x})
\label{S_p},
\end{align}
while the spin-blocked Fock matrices may be written as
\begin{align}
F^{\elab,\sa\sb}_{\eaoa\eaob} &= h_{\eaoa\eaob}^{\elab,\sa\sb} + G^{\eelab,\sa\sb}_{\eaoa\eaob}(\mathbf{D}^{\elab}) + 
G^{\eplab,\sa\sb}_{\eaoa\eaob}(\mathbf{D}^{\plab}) +
{}^{\mathrm Z}F^{\elab,\sa\sb}_{\eaoa\eaob},
\label{F_e} \\
F^{\plab,\sa\sb}_{\paoa\paob} &= h_{\paoa\paob}^{\plab,\sa\sb} + G^{\pplab,\sa\sb}_{\paoa\paob}(\mathbf{D}^{\plab}) + 
G^{\pelab,\sa\sb}_{\paoa\paob}(\mathbf{D}^{\elab}) -
{}^{\mathrm Z}F^{\plab,\sa\sb}_{\paoa\paob},
\label{F_p}
\end{align}
in terms of the two-particle matrix elements
\begin{align}
G^{\eelab,\sa\sb}_{\eaoa\eaob}(\mathbf{D}^{\elab}) &= \sum_{\sc\sd}\sum_{\eaoc\eaod}D^{\elab,\sd\sc}_{\eaod\eaoc}(\eaosa\eaosb\Vert \eaosc\eaosd),
\label{G_ee} \\
G^{\pplab,\sa\sb}_{\paoa\paob}(\mathbf{D}^{\plab}) &= \sum_{\sc\sd}\sum_{\paoc\paod}D^{\plab,\sd\sc}_{\paod\paoc}(\paosa\paosb\Vert \paosc\paosd), 
\label{G_pp} \\
G^{\eplab,\sa\sb}_{\eaoa\eaob}(\mathbf{D}^{\plab}) &= \sum_{\sc\sd}\sum_{\paoc\paod}D^{\plab,\sd\sc}_{\paod\paoc}(\eaosa\eaosb\vert \paosc\paosd),
\label{G_ep} \\
G^{\pelab,\sa\sb}_{\paoa\paob}(\mathbf{D}^{\plab}) &= \sum_{\sc\sd}\sum_{\eaoc\eaod}D^{\elab,\sd\sc}_{\eaod\eaoc}(\paosa\paosb\vert \eaosc\eaosd),
\label{G_pe}
\end{align}
and spin-Zeeman matrix elements
\begin{align}
{}^{\mathrm Z}\mathbf{F}^{\elab} = \frac{1}{2m^\elab} \begin{pmatrix}
B_{z}\bm{\mathscr{S}^{\elab}} & (B_x - \ilab B_y)\bm{\mathscr{S}^{\elab}} \\
(B_x + \ilab B_y)\bm{\mathscr{S}^{\elab}} & -B_{z}\bm{\mathscr{S}^{\elab}}
\end{pmatrix}, \\
{}^{\mathrm Z}\mathbf{F}^{\plab} =  \frac{1}{2m^\plab} \begin{pmatrix}
B_{z}\bm{\mathscr{S}^{\plab}} & (B_x - \ilab B_y)\bm{\mathscr{S}^{\plab}} \\
(B_x + \ilab B_y)\bm{\mathscr{S}^{\plab}} & -B_{z}\bm{\mathscr{S}^{\plab}}
\end{pmatrix}.
\end{align}
The NEO-GHF equations, along with the NEO-UHF and NEO-RHF special cases, have been implemented in LONDON. The use of London orbitals ensures gauge-origin invariant calculations for molecular properties, where primitive, unnormalized electronic and protonic London orbitals, respectively, are given by
\begin{align}
\phi_{\eaoa}^{\elab}(\mathbf{r}^{\elab},\mathbf{R}) = \omega_{\eaoa}^{\elab}(\mathbf{r}^{\elab},\mathbf{R})\mathrm e^{\ilab q^{\elab}\mathbf{A}(\mathbf{R})\cdot\mathbf{r}^{\elab}}, \label{london_e}\\
\phi_{\paoa}^{\plab}(\mathbf{r}^{\plab},\mathbf{R}) = \omega_{\paoa}^{\plab}(\mathbf{r}^{\plab},\mathbf{R}) \mathrm e^{\ilab q^{\plab}\mathbf{A}(\mathbf{R})\cdot\mathbf{r}^{\plab}}, \label{london_p}
\end{align}
where $\omega^{\elab}_\eaoa(\mathbf{r}^{\elab},\mathbf{R})$ and $\omega^{\plab}_{\paoa}(\mathbf{r}^{\plab},\mathbf{R})$ are standard primitive Cartesian Gaussian functions.
Note the presence of the electronic and protonic charges in Eqs.~\eqref{london_e} and~\eqref{london_p}.

\subsection{Small oscillations in Hartree--Fock theory}

Here, we present a derivation of TDHF theory, also known as the random phase approximation (RPA), for the generally complex case, which is applicable in the presence of a magnetic field. The electronic TDHF equations have been derived in various ways, emphasizing either dynamical aspects~\cite{MCWEENY92,OLSEN_JCP82_3235,CHRISTIANSEN_IJQC68_1} or that they yield approximations to static excitation energies~\cite{ALTICK_PR133_632,DUNNING_JCP47_1735} (see also recent work explicitly considering magnetic fields~\cite{Sen2019}). In the present NEO framework, we provide a derivation that emphasizes the analogy with the semiclassical vibrational problem and is valid in the presence of a magnetic field. The NEO-TDHF equations have been previously derived for the purely real-valued case in the absence of a magnetic field.\cite{Yang2018-1} 

 \subsubsection{TDHF equations in the time and frequency domains}

We determine the time development of the NEO Hartree--Fock wave function by the Dirac--Frenkel variation principle. For a general time-dependent 
wave function $\psi(t)$ and Hamiltonian $H$, the Dirac--Frenkel action functional is given by~\cite{BROECKHOVE_CPL149_547}
\begin{equation}
S(\psi)  = \int_{t_0}^{t_1}\!\!L(\psi,\dot{\psi},t) \,\mathrm d t .
\end{equation}
where the Lagrangian is
\begin{equation}
L(\psi,\dot{\psi},t)  =  \bra{\psi(t)} H - \ilab\partial_t   \ket{\psi(t)}.
\end{equation}
The evolution of the wave function from time $t_0$ to $t_1$ is determined by making $S(\Psi)$ stationary with
respect to all variations in $\Psi$ consistent
with fixed values at the end points.
We parameterize the wave function as
\begin{equation}
\ket{\psi(t)} = \mathrm e^{\ilab\kappa(t)} \ket{0},
\end{equation}
where $\kappa(t)$ is the time-dependent Hermitian orbital-rotation operator given by
\begin{align}
  \kappa(t) & = \sum_{\genoa\genob} \kappa_{\genoa\genob}(t) \genexc_{\genoa\genob},
  \quad \genexc_{\genoa\genob} =c^{\dagger}_{\genoa} c_{\genob},
\end{align}
and $\vert 0 \rangle$ is a time-independent reference wave function, which is taken to satisfy the conditions
\begin{equation}
\langle 0 \vert [ H, \genexc_{\genoa\genob}] \vert 0 \rangle = 0
\label{Hk0}
\end{equation}
for each pair $\genoa,\genob$.
With this parameterization,
the integrand in the Dirac--Frenkel action functional becomes
\begin{equation}
L(\psi,\dot{\psi},t) 
= \langle 0 \vert \mathrm e^{-\mathrm i \kappa(t)}
(H - \mathrm i \partial_t) \mathrm e^{\mathrm i \kappa(t)}\vert 0 \rangle.
\end{equation}
Performing a Baker--Campbell--Hausdorff (BCH) expansion and noting that $[- \mathrm i \partial_t, \mathrm i \kappa] = \dot \kappa$, 
we arrive at the following expression for the action integrand in terms of $\kappa(t)$ and $\dot \kappa(t)$:
\begin{align}
L(\kappa(t),\dot \kappa(t)) &= \left\langle 0 \left\vert H - \mathrm i \partial_t \right\vert 0 \right\rangle \nonumber \\ &\quad
+ \mathrm i\left\langle 0 \left\vert [H,  \kappa(t)] - \mathrm i \dot \kappa(t) \right\vert 0 \right\rangle 
 \nonumber \\
 &  \quad\quad  - 
\tfrac{1}{2}\left\langle 0 \left\vert [[H,   \kappa(t)]- \mathrm i \dot \kappa(t), \kappa(t)] \right\vert 0 \right\rangle \nonumber \\ &\quad\quad\quad+ \mathcal O(\kappa^3),
\label{bch}
\end{align}
where
$\langle 0 \vert -\mathrm i \partial_t \vert 0 \rangle = 0$
since $\vert 0 \rangle$ is time independent and 
$\langle 0 \vert [H,  \kappa] \vert 0 \rangle = 0$
by Eq.\,\eqref{Hk0}.
From the requirement that
\begin{equation}
\delta S = \int_{t_0}^{t_1} \! \delta L(\kappa(t),\dot\kappa(t)) \, \mathrm d t = 0
\end{equation}
 subject to $\delta \kappa_{\genoa\genob}(t_0)= 0$ and $\delta \kappa_{\genoa\genob}(t_1)= 0$ for each $\delta \kappa_{\genoa\genob}(t)$, we obtain the variational conditions
\begin{equation}
\frac{\partial L(\kappa(t),\dot \kappa(t))}{\partial\kappa_{\genoa\genob}(t)} =
\frac{\mathrm d }{\mathrm d t} \frac{\partial L(\kappa(t),\dot \kappa(t))}{\partial \dot \kappa_{\genoa\genob}(t)} .
\end{equation}
Inserting the BCH expression in Eq.\,\eqref{bch}, we obtain for each pair $\genoa,\genob$
\begin{align}
  \ilab \bra{0} [\dot{\kappa}(t), \genexc_{\genoa\genob}] \ket{0} &=  \tfrac{1}{2} \left\langle 0\left\vert [[H,\kappa(t)], \genexc_{\genoa\genob}] \right \vert 0 \right\rangle 
  \nonumber \\ 
  & \quad + \tfrac{1}{2}\left\langle 0\left\vert  [[H, \genexc_{\genoa\genob}],\kappa(t)] \right \vert 0 \right\rangle \nonumber \\ &\quad\quad+ 
  \mathcal O(\kappa^2) .\label{eq53}
\end{align}
Using the commutator identity $[A,[B,C]] + [B,[C,A]] + [C,[A,B]] = 0$
and noting that $\bra{0} [[\genexc_{\genoa\genob}, \kappa(t)],H] \ket{0}=0$ by Eq.\,\eqref{Hk0} since
$[\tau_{\genoa\genob}, \kappa(t)]$ is
a linear combination of excitation operators, we find that
\begin{align}
  \ilab \bra{0} [\dot{\kappa}(t), \tau_{\genoa\genob}] \ket{0} &=  \left\langle 0\left\vert [[H,\kappa(t)], \tau_{\genoa\genob}] \right \vert 0 \right\rangle \nonumber \\ & \qquad+
  \mathcal O(\kappa^2(t)),
  \label{eq54}
\end{align}
 which upon truncation at second order in $\kappa(t)$ gives the TDHF equations in the time domain. Finally, transforming to the frequency domain
using the Fourier transform convention in Eq.\,\eqref{eqFourierConvention}, we arrive at
the standard TDHF equations
\begin{align}
\omega \bra{0} [\genexc_{\genoa\genob},\hat{\kappa}(\omega)] \ket{0} = \bra{0} [\genexc_{\genoa\genob}, [ H, \hat{\kappa}(\omega)]]\ket{0},
\label{rpa_002}
\end{align}
for each pair $\genoa,\genob$.

 \subsubsection{NEO-TDHF equations}

Before evaluating Eq.~\eqref{rpa_002}, we note that the $\kappa(t)$ operator is restricted to contain only particle-conserving excitation and de-excitation operators. Additionally, only occupied--virtual excitations and de-excitations are considered. Finally, Hermiticity of $\kappa(t)$ demands that
\begin{align}
  \kappa(t) &= \sum_{ia}\kappa^{\elab}_{ai}(t)a^\dagger_a a_i + \sum_{ia}\kappa^{\elab *}_{ai}(t)a^\dagger_i a_a \nonumber \\
  &\quad + \sum_{IA}\kappa^{\plab}_{AI}(t)b^\dagger_A b_I + \sum_{IA}\kappa^{\plab *}_{AI}(t)b^\dagger_I b_A,
  \label{rpa_003}
\end{align}
where we treat $\kappa^{\elab}_{ai}(t) \ \text{and} \ \kappa^{\elab *}_{ai}(t)$ and likewise
$\kappa^{\plab}_{AI}(t)$ and $\kappa^{\plab *}_{AI}(t)$ as independent parameters. We may then write the Fourier transforms $\mathscr{F}$ of these functions as
\begin{align}
    \mathscr{F}[\kappa^{\elab}_{ai}(t)] = \hat{\kappa}^{\elab}_{ai}(\omega) = X^{\elab}_{ai}(\omega), \label{rpa_004} \\
    \mathscr{F}[\kappa^{\elab *}_{ai}(t)] = [\hat{\kappa}^{\elab}_{ai}(-\omega)]^{*} = [X^{\elab}_{ai}(-\omega)]^{*} = -Y^{\elab}_{ai}(\omega), \label{rpa_005}
\end{align}
with analogous equations for the protonic operators. Under the stated constraints, there are four cases to examine:
\begin{align}
\omega \bra{0} [a^{\dagger}_i a_a,\hat{\kappa}(\omega)] \ket{0} = \bra{0} [a^{\dagger}_i a_a, [ H, \hat{\kappa}(\omega)]]\ket{0}, \label{rpa_006} \\
\omega \bra{0} [a^{\dagger}_a a_i,\hat{\kappa}(\omega)] \ket{0} = \bra{0} [a^{\dagger}_a a_i, [ H, \hat{\kappa}(\omega)]]\ket{0}, \label{rpa_007} \\
\omega \bra{0} [b^{\dagger}_I b_A,\hat{\kappa}(\omega)] \ket{0} = \bra{0} [b^{\dagger}_I b_A, [ H, \hat{\kappa}(\omega)]]\ket{0}, \label{rpa_008} \\
\omega \bra{0} [b^{\dagger}_A b_I,\hat{\kappa}(\omega)] \ket{0} = \bra{0} [b^{\dagger}_A b_I, [ H, \hat{\kappa}(\omega)]]\ket{0}. \label{rpa_009}
\end{align}
Beginning with Eq.~\eqref{rpa_006}, we have
\begin{align}
&A^{\elab}_{ai,bj}X^{\elab}_{bj} + B^{\elab}_{ai,bj}Y^{\elab}_{bj} \nonumber \\
&\quad\quad + T^{\elab}_{ai,AI}X^{\plab}_{AI} + R^{\elab}_{ai,AI}Y^{\plab}_{AI} = \omega X^{\elab}_{bj},
\label{rpa_010}
\end{align}
where
\begin{align}
    A^{\elab}_{ai,bj} &= \bra{0} [a^{\dagger}_i a_a, [ H, a^\dagger_b a_j]]\ket{0} \nonumber \\
    &= F^{\elab}_{ab}\delta_{ij} - F^{\elab}_{ji}\delta_{ab} + (ai\Vert jb),\label{rpa_011} \\
    B^{\elab}_{ai,bj} &= -\bra{0} [a^{\dagger}_i a_a, [ H, a^\dagger_j a_b]]\ket{0} = (ai\Vert bj),\label{rpa_012} \\
    T^{\elab}_{ai,AI} &= \bra{0} [a^{\dagger}_i a_a, [ H, b^\dagger_A b_I]]\ket{0}=-(ai\vert IA),\label{rpa_013}\\
    R^{\elab}_{ai,AI} &= -\bra{0} [a^{\dagger}_i a_a, [ H, b^\dagger_I b_A]]\ket{0}=-(ai\vert AI)\label{rpa_014},
\end{align}
while Eq.~\eqref{rpa_008} gives the same results, mutatis mutandis, for the protonic case:
\begin{align}
&A^{\plab}_{AI,BJ}X^{\plab}_{BJ} + B^{\plab}_{AI,BJ}Y^{\plab}_{BJ} \nonumber \\
&\quad\quad + T^{\plab}_{AI,ai}X^{\elab}_{ai} + R^{\plab}_{AI,ai}Y^{\elab}_{ai} = \omega X^{\plab}_{BJ},
\label{rpa_015}
\end{align}
with
\begin{align}
    A^{\plab}_{AI,BJ} &= \bra{0} [b^{\dagger}_I b_A, [ H, b^\dagger_B b_J]]\ket{0} \nonumber \\
    &= F^{\plab}_{AB}\delta_{IJ} - F^{\plab}_{JI}\delta_{AB} + (AI\Vert JB),\label{rpa_016} \\
    B^{\plab}_{AI,BJ} &= -\bra{0} [b^{\dagger}_I b_A, [ H, b^\dagger_J b_B]]\ket{0} = (AI\Vert BJ),\label{rpa_017}\\
    T^{\plab}_{AI,ai} &= \bra{0} [b^{\dagger}_I b_A, [ H, a^\dagger_a a_i]]\ket{0} = -(AI\vert ia),\label{rpa_018}\\
    R^{\plab}_{AI,ai} &= -\bra{0} [b^{\dagger}_I b_A, [ H, a^\dagger_i a_a]]\ket{0} = -(AI\vert ai) \label{rpa_019}.
\end{align}
At this stage, we note that Eq.~\eqref{rpa_007} is equivalent to the conjugate of Eq.~\eqref{rpa_006} followed by the substitution $\omega \rightarrow -\omega$, with the same relationship holding for Eqs.~\eqref{rpa_008} and~\eqref{rpa_009}. Additionally, it is clear that $\mathbf{T}^{\elab} = \mathbf{T}^{\plab\dagger}$ and $\mathbf{R}^{\elab} = \mathbf{R}^{\plab\text{T}}$, so we need only specify generic  ``\textbf{T}" and ``\textbf{R}" matrices
\begin{align}
    T_{ai,AI} = -(ai\vert IA),\label{rpa_020} \\
    R_{ai,AI} =-(ai\vert AI).\label{rpa_021}
\end{align}
Thus Eqs.~\eqref{rpa_006} --~\eqref{rpa_009} constitute a generalized eigenvalue problem
\begin{align}
&\begin{pmatrix}
\mathbf{A}^{\elab} & \mathbf{B}^{\elab} & \mathbf{T} & \mathbf{R} \\
\mathbf{B}^{\elab *} & \mathbf{A}^{\elab *} & \mathbf{R}^{*} & \mathbf{T}^{*} \\
\mathbf{T}^{\dagger} & \mathbf{R}^{\text{T}} & \mathbf{A}^{\plab} & \mathbf{B}^{\plab} \\
\mathbf{R}^{\dagger} & \mathbf{T}^{\text{T}} & \mathbf{B}^{\plab *} & \mathbf{A}^{\plab *} \\
\end{pmatrix} 
\begin{pmatrix}
\mathbf{X}^{\elab}  \\
\mathbf{Y}^{\elab} \\
\mathbf{X}^{\plab}  \\
\mathbf{Y}^{\plab} \\
\end{pmatrix} 
 \nonumber \\ & \qquad \quad \qquad \qquad =\omega
\begin{pmatrix}
\mathbf{I} & \mathbf{0} & \mathbf{0} & \mathbf{0} \\
\mathbf{0} & -\mathbf{I} & \mathbf{0} & \mathbf{0} \\
\mathbf{0} & \mathbf{0} & \mathbf{I} & \mathbf{0} \\
\mathbf{0} & \mathbf{0} & \mathbf{0} & -\mathbf{I} \\
\end{pmatrix} 
\begin{pmatrix}
\mathbf{X}^{\elab}  \\
\mathbf{Y}^{\elab} \\
\mathbf{X}^{\plab}  \\
\mathbf{Y}^{\plab} \\
\end{pmatrix}.
\label{rpa_022}
\end{align}
In Eq.~\eqref{rpa_022}, $\mathbf{X}^{\elab}$ and $\mathbf{X}^{\plab}$ are the electronic and protonic excitation amplitudes, respectively, $\mathbf{Y}^{\elab}$ and $\mathbf{Y}^{\plab}$ are the electronic and protonic de-excitation amplitudes, respectively, and $\omega$ are the eigenvalues, which may be of protonic, electronic, or mixed character. The close analogy with the semiclassical treatment makes it reasonable to rely on spatial symmetry to match quantum modes $\hat{\kappa}(\omega)$ that are dominated by the nuclear part to their semiclassical counterparts. Care is required, however, as the dimensionality of the TDHF and semiclassical problems differ and the former typically yields many more modes even when the electronic component is disregarded. 

The UHF and RHF variants of Eq.~\eqref{rpa_022} are generated through appropriate restrictions on spin. For the purposes of this work, we are interested in an electronic RHF reference with a single quantum proton. In this specific case, spin-adaptation of Eq.~\eqref{rpa_022} reveals that only the electronic singlets couple with the protonic excitations. Summation over electronic spin and  the variable substitutions $\mathbf{X}^{\plab} \rightarrow \sqrt{2}\mathbf{X}^{\plab}, \mathbf{Y}^{\plab} \rightarrow \sqrt{2}\mathbf{Y}^{\plab}$ yield
\begin{align}
&\begin{pmatrix}
\rpamat{A}{\elab,\sigma}{} & \rpamat{B}{\elab,\sigma}{} & \sqrt{2}\rpamat{T}{}{} & \sqrt{2}\rpamat{R}{}{} \\
\rpamat{B}{\elab,\sigma*}{} & \rpamat{A}{\elab,\sigma*}{} & \sqrt{2}\rpamat{R}{*}{} & \sqrt{2}\rpamat{T}{*}{} \\
\sqrt{2}\rpamat{T}{\dagger}{} & \sqrt{2}\rpamat{R}{\text{T}}{} & \rpamat{A}{\plab}{} & \rpamat{B}{\plab}{} \\
\sqrt{2}\rpamat{R}{\dagger}{} & \sqrt{2}\rpamat{T}{\text{T}}{} & \rpamat{B}{\plab*}{} & \rpamat{A}{\plab*}{} \\
\end{pmatrix} 
\begin{pmatrix}
\rpamat{X}{\elab}{\uparrow}+\rpamat{X}{\elab}{\downarrow}  \\
\rpamat{Y}{\elab}{\uparrow}+\rpamat{Y}{\elab}{\downarrow} \\
\sqrt{2}\rpamat{X}{\plab}{\uparrow} \\
\sqrt{2}\rpamat{Y}{\plab}{\uparrow} \\
\end{pmatrix} 
\nonumber \\
&\qquad\ = \omega
\begin{pmatrix}
\mathbf{I} & \mathbf{0} & \mathbf{0} & \mathbf{0} \\
\mathbf{0} & -\mathbf{I} & \mathbf{0} & \mathbf{0} \\
\mathbf{0} & \mathbf{0} & \mathbf{I} & \mathbf{0} \\
\mathbf{0} & \mathbf{0} & \mathbf{0} & -\mathbf{I} \\
\end{pmatrix}
\begin{pmatrix}
\rpamat{X}{\elab}{\uparrow}+\rpamat{X}{\elab}{\downarrow}  \\
\rpamat{Y}{\elab}{\uparrow}+\rpamat{Y}{\elab}{\downarrow} \\
\sqrt{2}\rpamat{X}{\plab}{\uparrow} \\
\sqrt{2}\rpamat{Y}{\plab}{\uparrow} \\
\end{pmatrix},
\label{rpa_023}
\end{align}
where
\begin{align}
    A^{\elab,\sigma}_{ai,bj} &= (\epsilon_a - \epsilon_i)\delta_{ij}\delta_{ab} + 2(ai\vert jb) - (ab\vert ji), \label{rpa_024} \\
    B^{\elab,\sigma}_{ai,bj} &= 2(ai\vert bj) - (aj\vert bi), \label{rpa_025}
\end{align}
are the spin-adapted singlet variants of their GHF counterparts, written in terms of spin-free orbitals, and $\rpamat{X}{\elab}{\uparrow} \ (\rpamat{X}{\elab}{\downarrow})$ refers to the spin up (down) excitation amplitudes, with the de-excitations defined analogously. Note that the other blocks $\mathbf{T}, \ \mathbf{R}, \ \mathbf{A}^{\plab}, \ \text{and} \ \mathbf{B}^{\plab}$ appearing in Eq.~\eqref{rpa_023} are identical in form to their GHF counterparts (see Eqs.\eqref{rpa_013},~\eqref{rpa_014},~\eqref{rpa_016}, \ and~\eqref{rpa_017}), with the salient difference being that in the GHF case the orbitals in question are spinors, while in the spin-adapted case the orbitals are spin-free. 

Finally, as a brief digression on the use of language, we point out that terms such as ``mode" are only technically appropriate in their application to the QEP. In NEO-TDHF theory, there are no classical ``modes". In Section \ref{exc_char_sec}, we will demonstrate a scheme for the association of the quantum excitations to their semi-classical counterparts. However, it is to be understood that terms like ``stretching mode" or ``precession" in the quantum-mechanical context are used only by analogy to the classical case, and do not suggest that the NEO-TDHF case itself is  classical.

\subsection{Characterization of NEO-TDHF excitations} \label{exc_char_sec}

In this work, Eq.~\eqref{rpa_023} is solved by inversion of the metric followed by diagonalization. The resulting excitations may be characterized as electronic or protonic according to the dominance of electronic vs.\ protonic amplitudes in the eigenvectors -- that is, $\mathbf{X}^{\plab} \approx \mathbf 0$ and $\mathbf{Y}^{\plab} \approx  \mathbf 0$ for an electronic transition, and $\mathbf{X}^{\elab} \approx \mathbf 0$ and $\mathbf{Y}^{\elab} \approx \mathbf 0$ for a protonic transition.\cite{Yang2018-1,Culpitt2019,Pavosevic2020} To associate quantum modes with semi-classical modes, we examine the time evolution of the expectation value of the proton position
\begin{align}
    \bra{\psi(t)} \mathbf{r}^\plab \ket{\psi(t)} &= \bra{0} \mathrm e^{-\ilab\kappa(t)} \, \mathbf{r}^\plab \, \mathrm e^{\ilab\kappa(t)}\ket{0} \nonumber \\
    &\approx \bra{0}\mathbf{r}^\plab\ket{0} + \bra{0}[\mathbf{r}^\plab,\ilab\kappa(t)]\ket{0},
    \label{rpa_c_001}
\end{align}
where we have truncated the BCH expansion at first order in agreement with the stated assumption that $\kappa(t)$ is small. We are interested in the resonances of the Fourier series, which in the TDHF paradigm are taken to be excitation energies. For a particular resonance frequency $\omega$, we then have
\begin{align}
    \kappa(t) &= \hat{\kappa}(\omega)\,\mathrm e^{-\ilab\omega t} + \hat{\kappa}(-\omega)\,\mathrm e^{\ilab\omega t} \nonumber \\
    &= \hat{\kappa}(\omega)\, \mathrm e^{-\ilab\omega t} + \hat{\kappa}(\omega)^{*} \mathrm e^{\ilab\omega t}.
    \label{rpa_c_002}
\end{align}
Thus
\begin{align}
    \bra{\psi(t)} \mathbf{r}^\plab \ket{\psi(t)} &=  \bra{0}\mathbf{r}^\plab\ket{0} + \ilab\bra{0}[\mathbf{r}^\plab,\hat{\kappa}(\omega)]\ket{0}\mathrm e^{-\ilab\omega t} + \text{c.c.},
    \label{rpa_c_003}
\end{align}
where $\bra{0}[\mathbf{r}^\plab,\hat{\kappa}(\omega)]\ket{0}$ is the transition dipole moment vector
\begin{align}
    \bra{0}[\mathbf{r}^\plab,\hat{\kappa}(\omega)]\ket{0} &=
     \sum_{AI}[\bra{I}\mathbf{r}^{\plab}\ket{A}X^{\plab}_{AI} + \bra{A}\mathbf{r}^{\plab}\ket{I}Y^{\plab}_{AI}].
    \label{rpa_c_004}
\end{align}
The Fourier component $\boldsymbol{\eta}_{\mathrm{TDHF}}(\omega) = \ilab\bra{0}[\mathbf{r}^\plab,\hat{\kappa}(\omega)]\ket{0}$ and the time evolution $\boldsymbol{\eta}_{\mathrm{TDHF}}(t) = \ilab\bra{0}[\mathbf{r}^\plab,\hat{\kappa}(\omega)]\ket{0} \mathrm  e^{-\ilab\omega t} + \text{c.c.}$ may be compared to the corresponding semiclassical quantities obtained from Eq.~\eqref{eqSemiClassQEP}. For the purposes of this work, we are interested in a single quantum proton. In systems where there are multiple quantum protons that can be taken as approximately distinguishable,\cite{Culpitt2019-2} the procedure outlined above will also be applicable. However, in the general case of multiple quantum protons, $\boldsymbol{\eta}_{\mathrm{TDHF}}(\omega)$ cannot be compared to the QEP, due to the indistinguishability of the protons. Other classification metrics will then be required, such as spatial symmetry, angular momentum, or density.

Following the same procedure as the expectation value of position, we can examine the time evolution of the density operator
\begin{align}
    n(\genvec) &= n^\elab(\genvec) + n^\plab(\genvec), \label{tot_den}
\end{align}
where
\begin{align}
    n^\elab(\genvec) &= \sum_{pq}\Phi^{\elab *}_p(\genvec)\Phi^{\elab}_q(\genvec) a^\dagger_p a_q, \label{eden} \\
    n^\plab(\genvec) &= \sum_{PQ}\Phi^{\plab *}_P(\genvec)\Phi^{\plab}_Q(\genvec) b^\dagger_P b_Q,
    \label{pden}
\end{align}
and
\begin{align}
        \bra{\psi(t)} n(\genvec) \ket{\psi(t)} &=  \bra{0}n(\genvec)\ket{0} 
        \nonumber \\ & \quad + \ilab\bra{0}[n(\genvec) ,\hat{\kappa}(\omega)]\ket{0}\mathrm e^{-\ilab\omega t} + \text{c.c.}.
    \label{rpa_c_006}
\end{align}
In Eq.~\eqref{rpa_c_006}, $\bra{0}[n(\genvec) ,\hat{\kappa}(\omega)]\ket{0}$ is the transition density
\begin{align}
    &\bra{0}[n(\genvec),\hat{\kappa}(\omega)]\ket{0} \nonumber \\
    &\quad = \sum_{ai}[\Phi^{\elab *}_i(\genvec)\Phi^{\elab}_a(\genvec)X^{\elab}_{ai} + \Phi^{\elab *}_a(\genvec)\Phi^{\elab}_i(\genvec)Y^{\elab}_{ai}] \nonumber \\
    &\quad\quad + \sum_{AI}[\Phi^{\plab *}_I(\genvec)\Phi^{\plab}_A(\genvec)X^{\plab}_{AI} + \Phi^{\plab *}_A(\genvec)\Phi^{\plab}_I(\genvec)Y^{\plab}_{AI}].
    \label{rpa_c_007}
\end{align}
Equations~\eqref{rpa_c_003} and~\eqref{rpa_c_004} along with Eqs.~\eqref{rpa_c_006} and~\eqref{rpa_c_007} allow for visualization of the time evolution of the protonic expectation value and total density, respectively, for a given frequency $\omega$. When an excitation is dominated by one type of particle, the other particle's contribution to Eq.~\eqref{rpa_c_007} can be ignored.

In previous work, transition-dipole-moment vectors as well as transition densities were used to interpret quantum modes calculated with NEO-TDHF theory.\cite{Yang2018-1,Yang2019,Culpitt2019,Culpitt2019-2} However, this was done only in the frequency domain, where transition-dipole-moment vectors from NEO calculations were compared with, for example, normal modes from a harmonic oscillator.\cite{Yang2019,Culpitt2019-2} Additionally, this was done in the absence of a magnetic field, where all quantities could be taken as real valued. 

In a magnetic field, the NEO transition-dipole-moment vectors and transition densities are complex valued. Consequently, it becomes more difficult to understand and interpret these quantities (and express them in the language of semi-classical modes) without recourse to visualization as function of time. Development of real-time NEO methods\cite{Zhao2020_RTneo} in a magnetic field may therefore be of interest for future investigations.

\section{Results}

In this section, we present vibrational excitation energies for the hydrogen nuclei of the HCN and FHF$^{-}$ molecules. Since these molecules have been thoroughly studied in the context of NEO-HF and NEO-TDHF theories in the absence of a magnetic field, they represent a good starting point for discussing similarities and differences between the field and field-free cases. The heavy nuclei (C, N, F) of each molecule are clamped, while the hydrogen nucleus and all electrons of each system are treated quantum mechanically with the NEO-TDHF method. Calculations are performed with the LONDON program, for a range of magnetic field strengths, starting from $B = 0$ and ending at $B = B_0 = 2.35 \times 10^5\,$T in $0.1B_0$ increments, with the field oriented parallel to the molecular axis in both cases. 

All calculations were performed with a singlet electronic NEO-RHF reference configuration and with the proton occupying a spin orbital oriented parallel to the field. The electronic basis set employed in all calculations is a decontracted cc-pVTZ  London basis (denoted Lu-cc-pVTZ) for the heavy nuclei and a  cc-pV5Z basis (denoted Lu-cc-pV5Z) for the hydrogen nuclei. The protonic basis set is a decontracted London orbital variant of the PB5-F basis set,\cite{Yu2020} denoted Lu-PB5-F. Note that the protonic basis set was developed without contraction.\cite{Yu2020} It has been shown that, for quantitative results, contracted cc-pV5Z and cc-pV6Z electronic basis sets are required for hydrogen nuclei treated with the NEO method,\cite{Yang2018-1,Culpitt2019} which is why we have chosen to use the decontracted cc-pV5Z basis. Molecular geometries were optimized at each field strength for the conventional electronic case, and the protonic basis functions were placed at the optimized nuclear positions (see Supplementary Material for molecular geometries). 

For each molecule, there are three protonic vibrational modes: two bending modes perpendicular to the molecular axis and a stretching mode along the molecular axis. The bending modes are degenerate in the absence of a magnetic field. However, once the field is introduced, the degeneracy of these modes is broken by the Lorentz force and the bending modes are then replaced by precessional modes about the  field axis -- a lower-frequency precession counterclockwise and a higher-frequency precession clockwise. The stretching modes are not split by the field since they lie directly along the field axis. Note that, in the complex-valued case, even zero-field degenerate bending modes can be represented as degenerate precessional modes by taking linear combinations. 

In the QEP calculations, the precessional frequencies are affected by the Berry screening force. We consider three cases. The first case, denoted QEP-sL (sL
= “screened Lorentz”), is the properly screened scenario, with contributions from both $\Omega^{\text{int}}$ and $\Omega^{\text{ext}}$ in Eq.~\eqref{omega_tot}. In the second case, denoted QEP-bL (bL = “bare Lorentz”), we set $\Omega^{\text{int}} = 0$ , which corresponds to the QEP with only the bare Lorentz force acting on the
nuclei. Finally, in the third case, denoted QEP-nL (nL
= “no Lorentz”), we set $\Omega^{\text{ext}} = 0$ and $\Omega^{\text{int}} = 0$, corresponding to no Lorentz force at all. These QEP cases are of interest to compare with the NEO-TDHF results because the NEO-TDHF frequencies should inherently contain the effects of $\Omega^{\text{tot}}$, including screening. Thus, we expect the NEO-TDHF calculations to match most closely with QEP-sL.

\begin{figure*}[h]
\centering
\begin{tabular}{ll}
(a) \\
\includegraphics[width=0.48\textwidth]{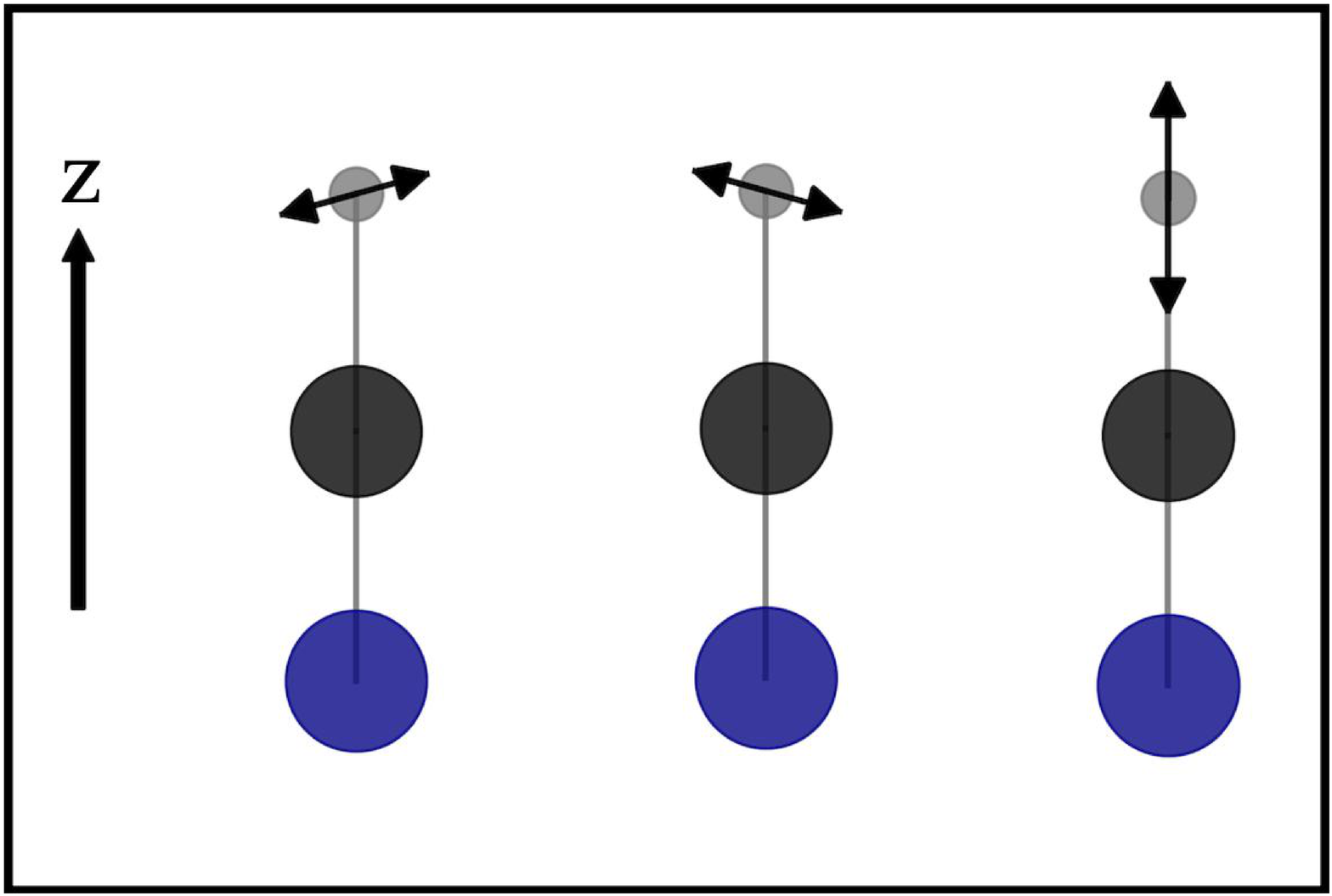} \\
(b) \\
\includegraphics[width=0.48\textwidth]{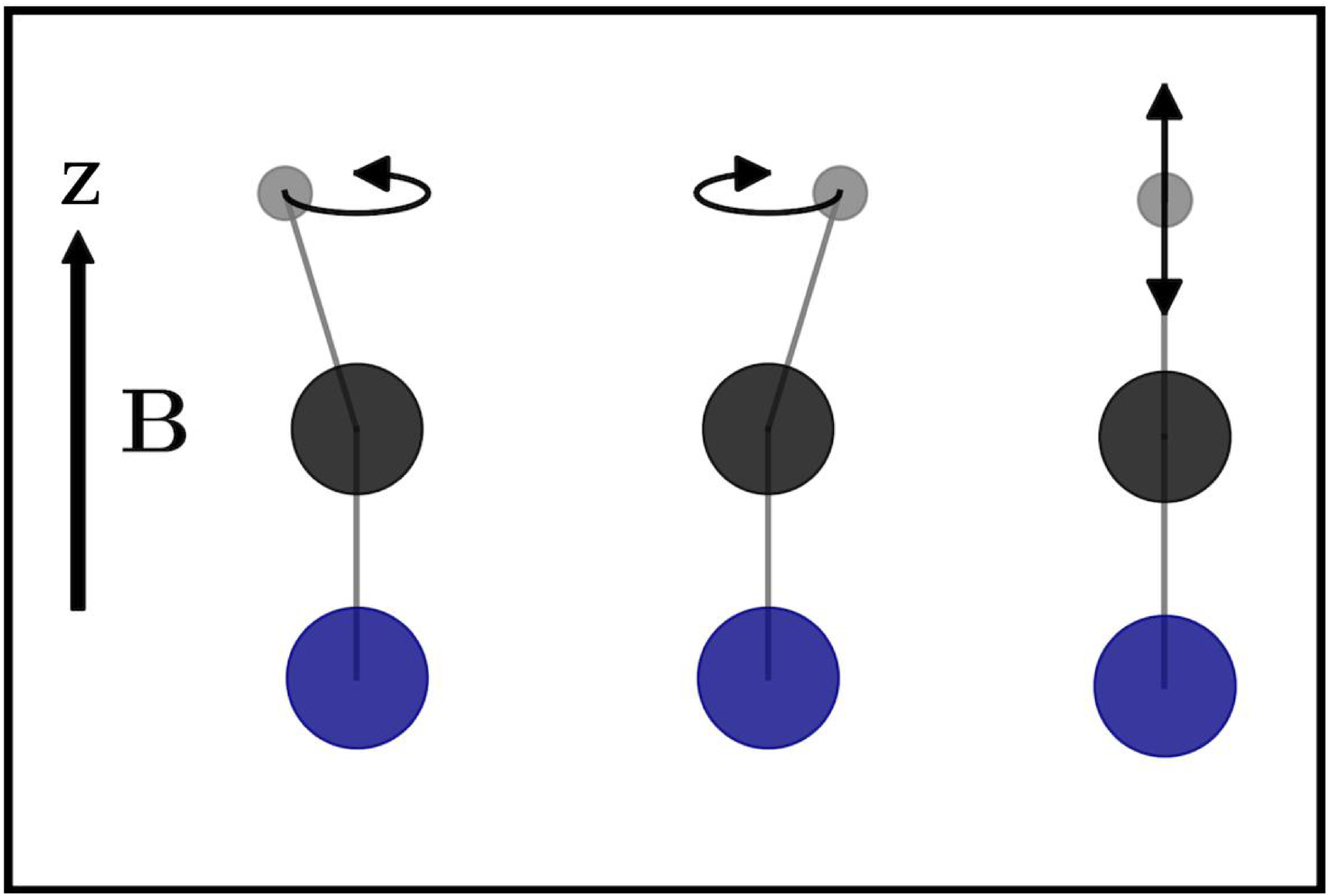} \\
\end{tabular}
\caption{Schematic representation of protonic vibrational modes of the HCN molecule oriented along the z-axis for clamped C and N nuclei in (a) zero magnetic field and (b) magnetic field oriented along the molecular axis. In panel (a) the two degenerate off-axis bending modes are replaced in panel (b) by non-degenerate precession modes. One precession is counterclockwise and the other is clockwise. The stretching mode is unaffected by the Lorentz force because its motion is parallel to the field. All depicted modes can be visualized as a function of time using the eigenvectors of the QEP  or the transition dipole moment vectors from NEO-TDHF theory as described in the text.}
\label{hcn_modes}
\end{figure*}

In Figure~\ref{hcn_modes}, the vibrational and precessonal modes of HCN are illustrated. In Tables~\ref{table_hcn} and~\ref{table_fhf}, we have listed the HCN and FHF$^-$ vibrational/precessional frequencies associated with the hydrogen atom for different magnetic field strengths, calculated with the NEO-TDHF and QEP methods. The QEP results are generated with a mass on the order of $10^{24}$ atomic units for the clamped nuclei.

When comparing NEO and QEP results, we are particularly interested in (1) the magnitudes of the vibrational excitation energies and (2) the magnitude of the splitting of the precessional modes in the screened and unscreened QEP cases. Regarding point (1), a quantitative comparison is not achievable, even in the BO limit, because the QEP neglects all anharmonic effects, which are relevant in both cases, especially so for the FHF$^-$ stretching mode. Nonetheless, qualitative comparisons can be made in the absence of large anharmonic effects. A better general benchmark would be, for example, a grid-based method, such as the Fourier grid Hamiltonian (FGH) method,\cite{Marston1989,Kurti1991,Webb2000} which has previously been used for benchmarking NEO-TDDFT.\cite{Yang2018-1,Culpitt2019} However, a grid-based machinery that fully incorporates the Lorentz force and Berry force is not currently available, although it could be developed in the future.

Regarding point (2), for the chosen orientation of the molecules to the field, it is demonstrated in Appendix~\ref{appendix_a} that the absolute value of the frequencies of precessional motion can be calculated according to the simple formula 
\begin{equation}
  |\omega_{\pm}| =   \sqrt{\frac{\omega_{\text{sc}}^2}{4}  + \omega_{\text{bend}}^2} \mp \frac{\omega_{\text{sc}}}{2}, \label{toy_bends_text}
\end{equation}
where the two cases correspond to clockwise ($\omega_{-}$) and counterclockwise ($\omega_{+}$) precessions about the field axis. 
Here, $\omega_{\mathrm{bend}}$ is the bending frequency as obtained from the PES in the harmonic approximation (QEP-nL), while $\omega_{\mathrm{sc}}$ is the screened cyclotron frequency $\omega_{\text{sc}} = (q^\plab B_z - \alpha) /m^\plab$, where $\alpha$ is an element of the Berry curvature. In the absence of screening, $\omega_{\text{sc}}$ reduces to the standard cyclotron frequency $\omega_{\text{c}} = q^\plab B_z /m^\plab$.

The calculated values of $|\omega_{\pm}|$ reproduce exactly the QEP-bL (using $\omega_\text c$) and QEP-sL (using $\omega_\text{sc}$) results. Reproduction of the splitting $|\omega_{-}|-|\omega_{+}| = \omega_{\text{sc}}$ is a particularly robust indicator of capturing the magnetic forces as the splitting only depends on cylindrical symmetry and is independent of $\omega_{\text{bend}}$ and the anharmonicty of the PES. This allows for a quantitative comparison of the description of magnetic effects at the (harmonic) QEP and (anharmonic) NEO levels, respectively. For a derivation of Eq.~\eqref{toy_bends_text}, see Appendix~\ref{appendix_a}.

\subsection{HCN}

\begin{figure*}[h]
\centering
\begin{tabular}{ll}
\includegraphics[width=0.48\textwidth]{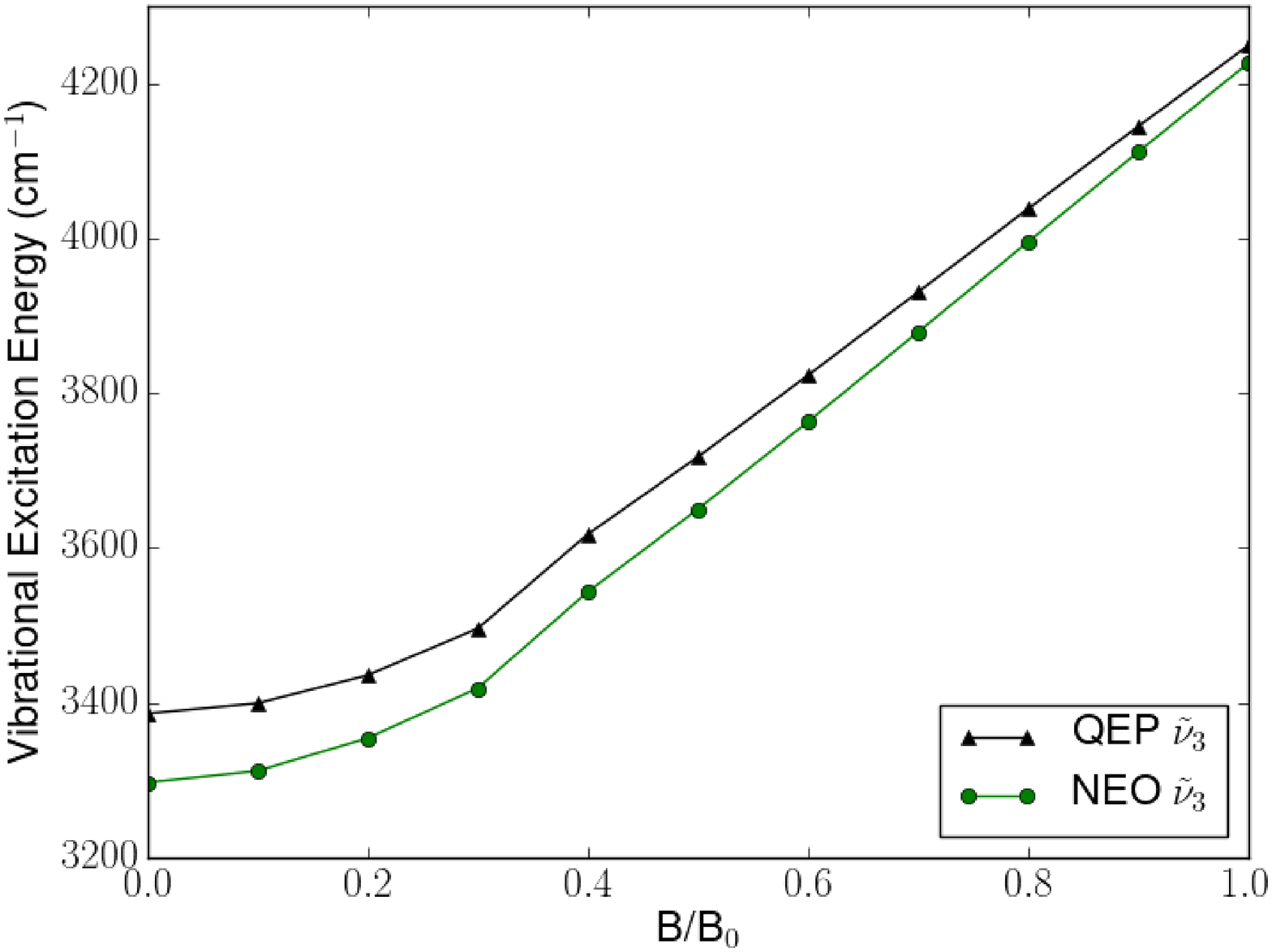} \\
\end{tabular}
\caption{Vibrational excitation energies corresponding to the stretching mode $\exclabc$ of HCN for the NEO-TDHF and QEP methods. Since the molecule is oriented parallel to the magnetic field, the QEP results are the same for all three variants (QEP-sL = QEP-bL = QEP-nL).}
\label{hcn_vib_stretch}
\end{figure*}

\begin{figure*}[h]
\centering
\begin{tabular}{ll}
\includegraphics[width=0.48\textwidth]{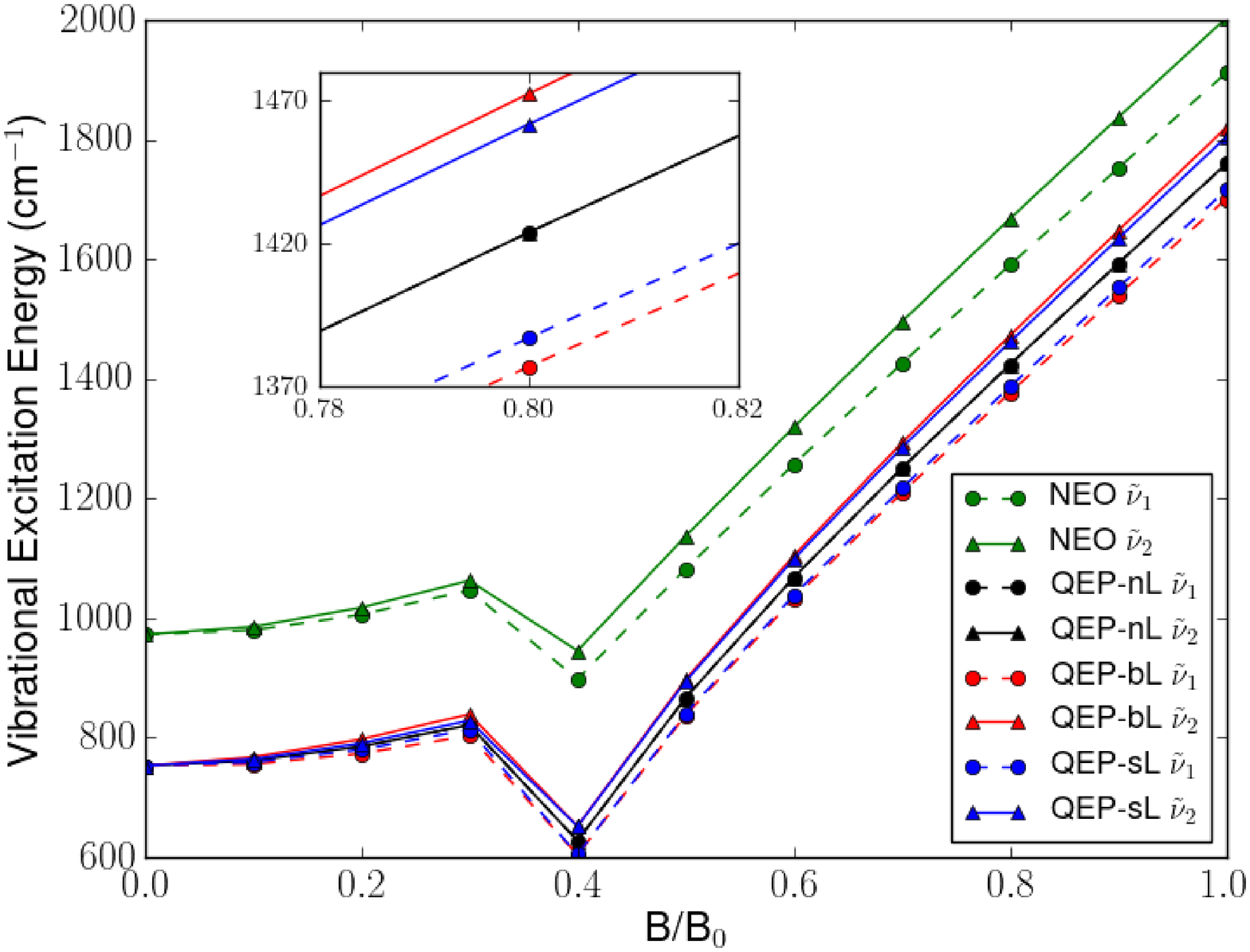} \\
\end{tabular}
\caption{Vibrational excitation energies corresponding to the precession modes $\exclaba$ and $\exclabb$ of HCN for the NEO-TDHF method and the three variants of the QEP method. Dashed lines with circles correspond to $\exclaba$ for a given method, while solid lines with triangles correspond to $\exclabb$. The inset highlights the different precessional splittings of the QEP-sL and QEP-bL methods. The discontinuity in the plot between $0.3B_0$ and $0.4B_0$ is due to a level crossing.}
\label{hcn_vib_bend}
\end{figure*}

Table~\ref{table_hcn} contains the  frequencies of the stretching and bending modes of HCN calculated using the NEO-TDHF and QEP methods (all three variants) as functions of the magnetic field strength. The stretching frequency is plotted in Fig.\,\ref{hcn_vib_stretch}, while Fig.\,\ref{hcn_vib_bend} shows the bending/precessional frequencies. The stretching frequencies are unaffected by the Lorentz force (by the parallel field orientation of the molecule) and are therefore the same for all QEP variants. We note the presence of a discontinuity in the plots, arising from a level crossing between $0.3B_0$ and $0.4B_0$. Examination of the molecular orbitals as well as the fact that the C-N bond at $0.4B_0$ is significantly longer than it is at $0.3B_0$ indicate that the C-N triple bond is reduced to a single bond and two nonbonding electron pairs at higher field strengths. By using the converged conventional electronic density as an initial guess for the NEO iterations, we were able to converge to the corresponding electronic state in the NEO calculations in this region.

Since HCN contains a terminal hydrogen, we would expect the NEO-TDHF stretching frequency to be lower than the corresponding QEP frequency due to anharmonicity. This is indeed observed in Fig.\,\ref{hcn_vib_stretch}, although the difference decreases with increasing field strength as the anharmonicity is affected by compression of the bond in the field. For the precessional modes, by contrast, the NEO frequencies are roughly 200\,cm$^{-1}$ higher than the QEP frequencies at all field strengths; see Fig.\,\ref{hcn_vib_bend}.

\begin{figure*}[h]
\centering
\begin{tabular}{ll}
\includegraphics[width=0.48\textwidth]{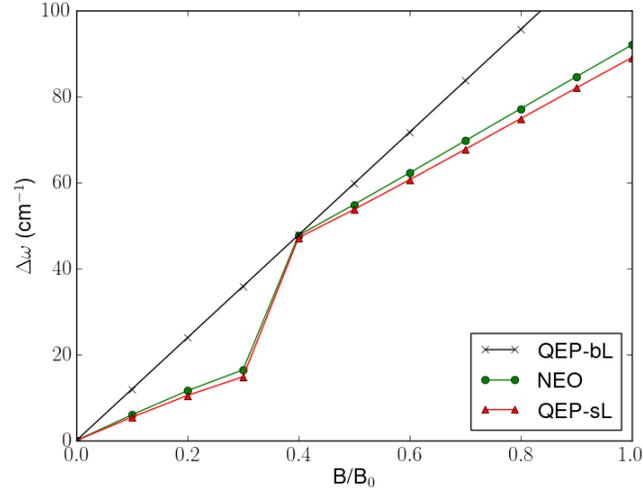} \\
\end{tabular}
\caption{Difference in energy between the two precessional modes of HCN for the NEO-TDHF, QEP-sL, and QEP-bL methods. For the QEP-bL method, the splitting is equal to the cyclotron frequency for the proton, ${q^\plab B_z}/{m^\plab}$. The discontinuity in the plot between $0.3B_0$ and $0.4B_0$ is due to a level crossing.}
\label{hcn_split}
\end{figure*}

An interesting observation in Fig.\,\ref{hcn_vib_bend} is that the QEP-bL and QEP-sL calculations do not give the same cyclotron splitting of the precessional modes, as highlighted in the inlay of the figure. This behaviour can be understood by noting that the QEP-bL splitting is always equal to the proton's cyclotron frequency, while the QEP-sL splitting vanishes in the limit of perfect screening. As seen in  Fig.\,\ref{hcn_split}, the NEO splitting agrees well with the QEP-sL results, suggesting that the NEO-TDHF method is capable of capturing the effects of both the bare Lorentz force and the Berry screening force. Note that the effect of the latter greatly decreases at field strengths greater than $0.3B_0$, again reflecting the different character of the ground state in this regime. However, the absence of screening at 0.4$B_0$, implied by the fact that the screened and unscreened splittings are nearly identical, might be an artifact of the level of theory in the proximity of an avoided crossing.

\subsection{FHF$^-$}

\begin{figure*}[h]
\centering
\begin{tabular}{ll}
\includegraphics[width=0.48\textwidth]{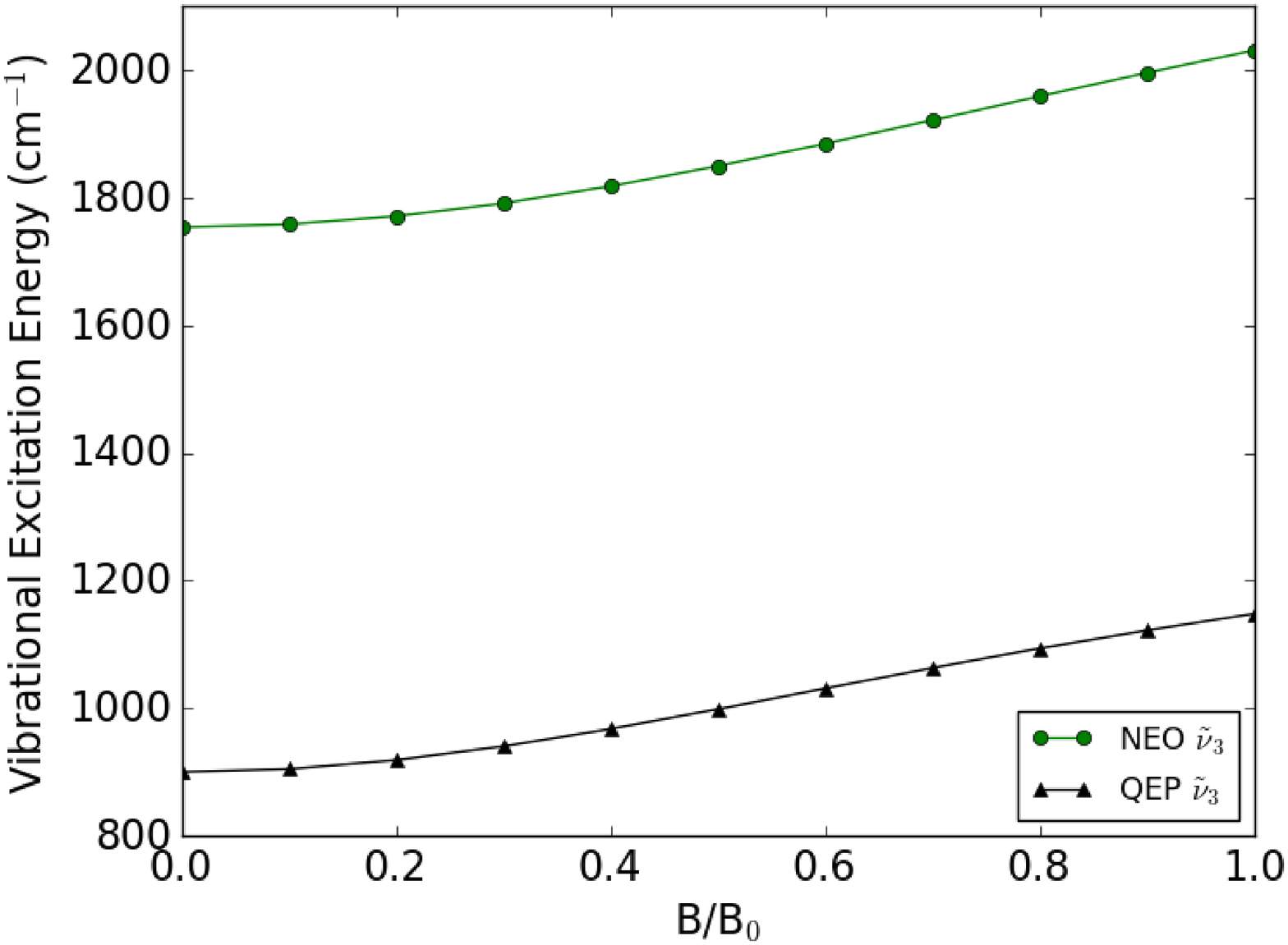} \\
\end{tabular}
\caption{Vibrational excitation energies corresponding to the stretching mode $\exclabc$ of FHF$^-$ for the NEO-TDHF and QEP methods. Since the molecule is oriented parallel to the magnetic field, the QEP results are the same for all three cases (QEP-sL = QEP-bL = QEP-nL).}
\label{fhf_vib_stretch}
\end{figure*}

\begin{figure*}[h]
\centering
\begin{tabular}{ll}
\includegraphics[width=0.48\textwidth]{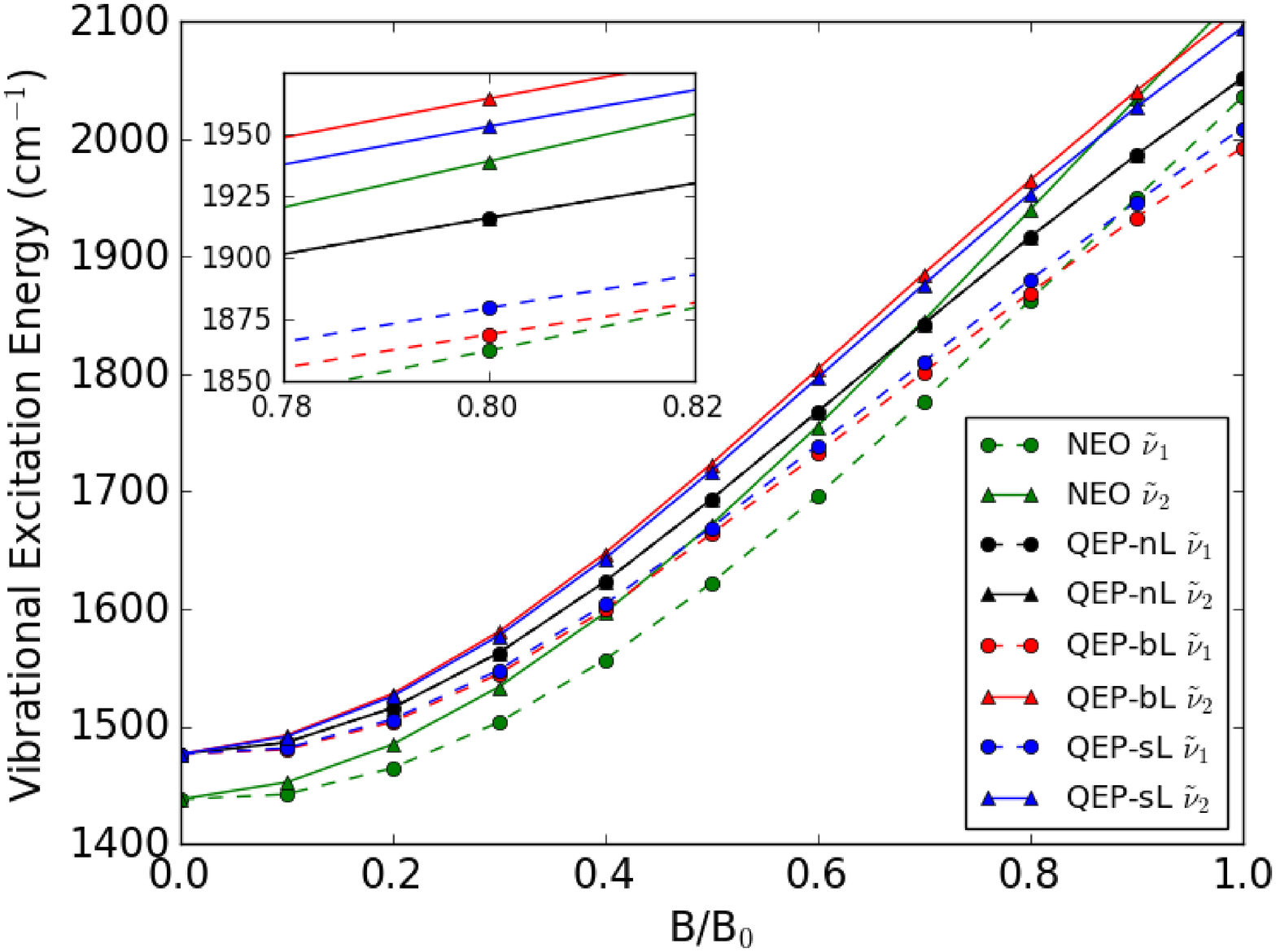} \\
\end{tabular}
\caption{Vibrational excitation energies corresponding to the precessional modes $\exclaba$ and $\exclabb$ of FHF$^-$ for the NEO-TDHF method and the three variants of the QEP method. Dashed lines with circles correspond to $\exclaba$ for a given method, while solid lines with triangles correspond to $\exclabb$. The inset highlights the different precessional splittings of the QEP-sL and QEP-bL methods.}
\label{fhf_vib_bend}
\end{figure*}

Table~\ref{table_fhf} contains the frequencies for the stretching and bending modes of FHF$^-$ 
calculated using the NEO and QEP methods as functions of the magnetic field strength. Figure\,\ref{fhf_vib_stretch} shows the stretching frequencies, while the bending/precessional frequencies are plotted in Fig.\,\ref{fhf_vib_bend}. The stretching frequencies are the same for all QEP variants, being unaffected by the Lorentz force due to parallel orientation of molecule to the field.

Previous zero-field studies using DFT and a grid-based FGH method have shown that the stretching frequency of FHF$^-$ is higher than the bending frequencies, in agreement with zero-field NEO-TDDFT results.\cite{Yang2018-1,Culpitt2019}  
Our field-dependent NEO-TDHF frequencies exhibit the same ordering except in the strongest fields, where the molecule is compressed to the point where one (at 0.9$B_0$) or both (at 1.0$B_0$) precessional modes become higher in energy than the stretching mode. By contrast, the QEP stretching modes are always lower in energy than the QEP precessional modes. The large difference in magnitude between the QEP  and NEO-TDHF stretching mode energies (see Fig.~\ref{fhf_vib_stretch}) is likely due to anharmonicity.

\begin{figure*}[h]
\centering
\begin{tabular}{ll}
\includegraphics[width=0.48\textwidth]{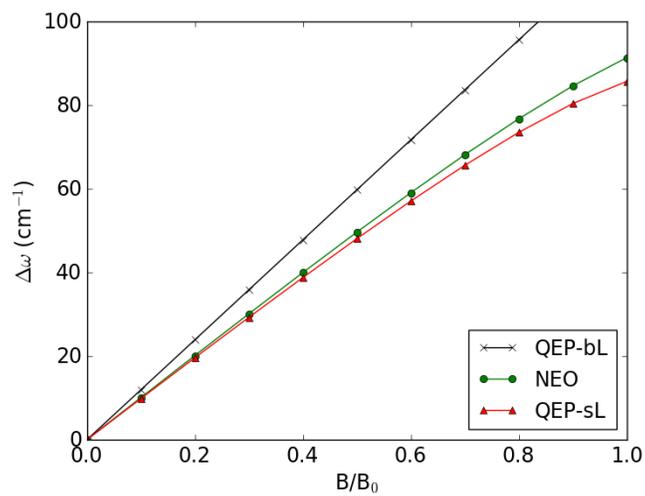} \\
\end{tabular}
\caption{Difference in energy between the two precessional modes of FHF$^-$ for the NEO-TDHF, QEP-sL, and QEP-bL methods. For QEP-bL method, the splitting is equal to the cyclotron frequency for the proton, ${q^\plab B_z}/{m^\plab}$.}
\label{fhf_split}
\end{figure*}

In FHF$^-$, the NEO and QEP-sL precessional frequencies are much closer to each other than in HCN, being always less than 100\,cm$^{-1}$ apart; see Fig.\,\ref{fhf_vib_bend}. As was the case for HCN, the cyclotron splitting of the precessional modes is reduced when screening is included by going from the QEP-bL level of theory to the QEP-sL level; see Fig.~\ref{fhf_split}. The NEO-TDHF results again agree very well with the QEP-sL results, demonstrating the capability of the NEO model to account for the effects of  screening  directly.

\section{Conclusions}

In this work, we have presented the theory and implementation of the NEO-GHF and NEO-TDHF models in the presence of a uniform magnetic field. The NEO-TDHF derivation proceeded through the Dirac--Frenkel action. For a given transition, the the protonic position expectation value and density as a function of time can be visualized, which aids in characterization of the excitations. The relevant equations were implemented in the LONDON program, which utilizes London orbitals for gauge-origin invariant calculations in a magnetic field. 

The NEO-TDHF results for the precessional and stretching modes of the HCN and FHF$^{-}$ molecules were compared to QEP results for three cases: QEP-nL (no Lorentz forces), QEP-bL (bare Lorentz forces, and QEP-sL (screened Lorentz forces). Because the QEP-sL model includes the screened Lorentz force, it is the most appropriate QEP variant for comparison with the NEO method. The NEO excitation energies were found to be within about 200\,cm$^{-1}$ of the QEP-sL energies except for the FHF$^-$ stretching mode, which showed a larger disparity, likely due to anharmonic effects. These differences are comparable to those observed in the zero-field case.

The degeneracy of the precessional (bending) mode is lifted in a  magnetic field by coupling to the cyclotron motion. The precessional splitting is independent of the PES in both the screened and bare Lorentz cases, allowing for quantitative comparisons of the NEO and QEP results. The QEP-bL splitting is always greater than the QEP-sL and NEO splittings, in agreement with the observation that degeneracy is recovered in the limit of perfect screening. The QEP-sL and NEO splittings are in excellent agreement, confirming that the NEO model accounts for the effects of the Berry curvature directly.

The  capability of NEO-TDHF to account for screening of the Lorentz force is significant, bearing in mind that the QEP-sL model requires the Berry curvature to be evaluated, making this model much more complex and expensive than the QEP-nL and QEP-bL models. By contrast, the NEO-TDHF model includes Lorentz screening automatically, in addition to non-BO and nuclear quantum effects. In our examples, these effects tend to reduce the screening experienced by hydrogen in the field with respect to the BO treatment. Therefore, future development of NEO methods in a magnetic field will be of interest, especially so for molecular dynamics, where screening is of great importance.

\section*{Supplementary Material}

See supplementary material for HCN and FHF$^-$ geometries.

\section*{Acknowledgements}

This work was supported by the Research Council of Norway through ‘‘Magnetic Chemistry’’ Grant No.\,287950 and CoE Hylleraas Centre for Quantum Molecular Sciences Grant No.\,262695. The work also received support from the UNINETT Sigma2, the National Infrastructure for High Performance Computing and Data Storage, through a grant of computer time (Grant No.\,NN4654K).

\section*{Data Availability}

The data that support the findings of this study are available within the article.

\appendix

\section{Bending vibration and precession for a cylindrically symmetric model system} \label{appendix_a}

\begin{figure*}[h]
\centering
\begin{tabular}{ll}
\includegraphics[width=0.48\textwidth]{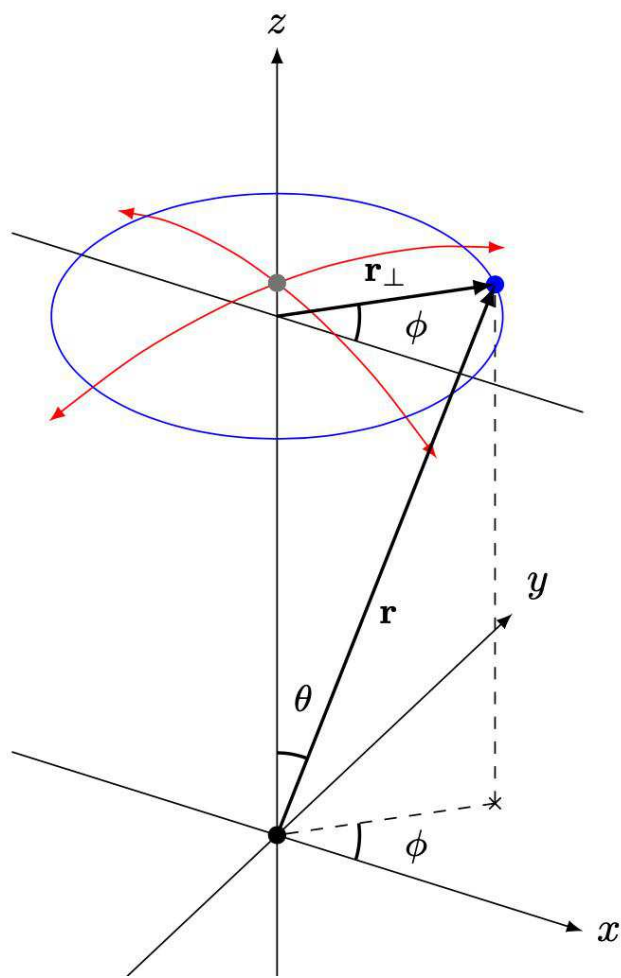} \\
\end{tabular}
\caption{Spherical coordinate system, bending modes in red, and precession in blue illustrated for a triatomic linear molecule.}
\label{appendix_model}
\end{figure*}

Consider a molecule with a linear equilibrium structure. We take the molecule to be aligned to the $z$-axis, with $N$ nuclear positions at $\mathbf{r}_j = (0,0,z_j)$,  labelled in order of decreasing coordinate $z_j > z_{j+1}$
with $j = 1,\ldots,N$.
Furthermore, we assume that the molecule has a bending mode that involves only the first atom, located at one end, the other nuclei being clamped. We fix the bond distance $R = |z_1-z_2|$ and assume cylindrical symmetry around the $z$-axis. We take into account the forces from the electronic PES as well as the bare Lorentz force and the Berry screening force. With the spherical coordinate system defined as in Fig.\,\ref{appendix_model}, the nuclear position and magnetic field are given by
\begin{align}
  \mathbf{r}_1 &= \begin{pmatrix} R\sin(\theta) \cos(\phi) \\ R \sin(\theta) \sin(\phi) \\ R \cos(\theta) \end{pmatrix}, \quad
  \mathbf{B} = \begin{pmatrix} 0 \\ 0 \\ B_z \end{pmatrix}.
\end{align}
The potential is taken to be a function $V(\theta)$ of $\theta$, while the magnetic forces are taken to be described by a vector potential that depends on $\theta$ in the manner
\begin{align}
\mathbf{a}_1(\theta) = \frac{1}{2}\mathbf{B}\times\mathbf{r}_1 = \frac{1}{2}B_z R \sin(\theta)\mathbf{e}_{\phi}.
\label{atheta}
\end{align}
Because we are solely interested in the motion of the nucleus at $\mathbf{r}_1$, we need only consider the screening force that arises from the Berry curvature at that nucleus (see Eq.~\eqref{newton_eqn}), corresponding to a single $I=J$ block of Eq.~\eqref{berry_curv}. The Berry screening force then becomes
\begin{align}
    \mathbf{\Omega}(\mathbf{r}) \dot{\mathbf{r}}_1  = \mathbf{s}_1(\mathbf{r})  \times \dot{\mathbf{r}}_1, \label{scr_force}
\end{align}
where $\mathbf{s}_1$ is the curl of the corresponding geometric vector potential 
\begin{align}
        \mathbf{s}_1(\mathbf{r}) = \boldsymbol \nabla \times \boldsymbol \chi_1(\mathbf{r}), \quad
   \boldsymbol{\chi}_1(\mathbf{r}) = \bra{\psi}{\mathbf p}_1\ket{\psi}. \label{gvp_def}
\end{align}
Here $\psi$ is the electronic wave function and ${\mathbf p}_1 = -\ilab\frac{\partial}{\partial \mathbf{r}_1}$ is the canonical momentum operator for the nucleus at $\mathbf{r}_1$. Note that $\mathbf{s}_1(\mathbf{r})$ and $\boldsymbol{\chi}_1(\mathbf{r})$ are functions of the coordinates of all nuclei because they depend on the electronic wave function {of the entire system}. In a molecule, the screening force generated by Eq.\,\eqref{scr_force} will therefore in general not be complete (i.e., exactly cancelling the bare Lorentz force), as it would for an atom undergoing similar motion.\cite{Culpitt2021}

For the given orientation of the molecule to the field, $\mathbf{s}_1(\mathbf{r})$ has only a nonzero $z$ component.\cite{Culpitt2021} We now make the additional assumption that, for small deviations of $\theta$ and $\phi$ around equilibrium, $\mathbf{s}_1(\mathbf{r})$ is  constant,
\begin{align}
  \mathbf{s}_1(\mathbf r) = \begin{pmatrix} 0 \\ 0 \\ \alpha \end{pmatrix}.
\end{align}
Under this assumption, we can represent the geometric vector potential in the symmetric gauge, analogous to the magnetic vector potential in Eq.\,\eqref{atheta}:
\begin{align}
\mathbf{\chi}_1(\theta) = \frac{1}{2}\mathbf{s}_1 \times \mathbf{r}_1 = \frac{1}{2}\alpha R \sin(\theta)\mathbf{e}_{\phi}.
\end{align}
The Lagrangian consists of one contribution identical to a three-dimensional pendulum, along with magnetic and geometric vector potentials 
\begin{equation}
  \begin{split}
    \mathcal{L} & = \frac{m}{2} \dot{\mathbf{r}}_1^2 + \dot{\mathbf{r}}_1\cdot [q \mathbf{a}_1(\theta) - \mathbf{\chi}_1(\theta)] - V(\theta)
    \\
    & = \frac{1}{2} m R^2 \dot{\theta}^2 + \frac{1}{2}m R^2 (\dot{\phi}^2 + \omega_{\text{sc}} \dot{\phi}) \sin(\theta)^2- V(\theta).
  \end{split}
\end{equation}
where the screened cyclotron frequency
\begin{equation}
\omega_{\text{sc}} = \frac{q B_z - \alpha}{m},
\label{omegasc}
\end{equation}
is calculated from the effective  magnetic field $B_z$ and a screening contribution $\alpha$. In the absence of screening, $\alpha=0$, we recover the standard cyclotron frequency $\omega_{\text{c}} = q B_z /m$. 
Calculating the generalized momenta, we obtain
\begin{align}
  p_{\theta} & = \frac{\partial \mathcal{L}}{\partial \dot{\theta}} = m R^2 \dot{\theta},
  \\
  p_{\phi} & = \frac{\partial \mathcal{L}}{\partial \dot{\phi}} = mR^2( \dot{\phi} + \tfrac{1}{2}  \omega_\text{sc}) \sin(\theta)^2,
\end{align}
while the generalized forces are given by
\begin{align}
  F_{\theta} & = \frac{\partial \mathcal{L}}{\partial \theta} = mR^2( \dot{\phi}^2 + \omega_\text{sc} \dot{\phi}) \sin(\theta) \cos(\theta) - \frac{\partial V}{\partial \theta},
  \\
  F_{\phi} & = \frac{\partial \mathcal{L}}{\partial \phi} = 0.
\end{align}
We will solve Lagrange's equations of motion  $\dot{p}_{\theta} = F_\theta$ and
$\dot{p}_{\phi} = F_\phi$ in two special cases: for a \emph{pure bending motion} with constant $\phi$ and varying $\theta$ and  for a \emph{pure precessional motion} with constant $\dot \phi$ and constant $\theta$. 

We consider the bending motion first. Since $F_\phi = 0$, the equation equation of motion for $\phi$ becomes $\dot{p}_{\phi} = 0$, implying that the momentum is conserved:
\begin{equation}
  p_{\phi} =m R^2 ( \dot{\phi} + \tfrac{1}{2} \omega_\text{sc}) \sin(\theta)^2 = C .
\end{equation}
A pure bending mode, with constant $\phi$ and varying $\theta$, is therefore possible only if $\omega_\text{sc} = 0$. From Eq.\,\eqref{omegasc},
we see that this happens either in a zero magnetic field where $q B_z = \alpha = 0$ or in a magnetic field with perfect screening $q B_z = \alpha$. In the harmonic approximation, $V(\theta) = \frac{1}{2} k \theta^2$, Lagrange's equation of motion $\dot{p}_{\theta} = F_{\theta}$  then simplifies to
\begin{equation}
  m R^2 \ddot{\theta} + k \theta = 0.
\end{equation}
Introducing the vibrational frequency
\begin{equation}
\omega_{\text{bend}} = \sqrt{\frac{k}{mR^2}},
\end{equation}
we may write the solution as $\theta(t) = \theta_0 \sin(\omega_{\text{bend}} t + \alpha_0)$ where $\alpha_0$ is a constant.

A pure precessional motion corresponds to constant $\dot{\phi}$ and constant $\theta$, and therefore $\dot{\theta} = 0$. The latter condition is possible if and only if $F_{\theta} = 0$ vanishes, yielding a quadratic equation for $\dot{\phi}$:
\begin{equation}
  \dot{\phi}^2 + \omega_\text{sc} \dot{\phi} - \frac{{\partial V}/{\partial \theta}}{m R^2 \sin(\theta) \cos(\theta)}= 0.
\end{equation}
In the harmonic approximation $V(\theta) \approx \tfrac{1}{2} k \theta^2$ with a sufficiently small value of $\theta$ so that $\sin(\theta) \approx \theta$ and $\cos(\theta) \approx 1$, we obtain the solutions
\begin{equation}
  \dot{\phi}_{\pm} \approx -\frac{\omega_{\text{sc}}}{2} \pm \sqrt{\frac{\omega_{\text{sc}}^2}{4}  + \omega_{\text{bend}}^2}. \label{toy_bends}
\end{equation}
Hence, the two possible precessional frequencies may be predicted from the effective cyclotron frequency (with or without screening) and the bending frequency. To demonstrate the accuracy of the outlined derivation, calculation of the magnitude of the two precessional frequencies according to Eq.\,\eqref{toy_bends} using only $\omega_{\text{bend}}$, $B_z$, and $\alpha$ as inputs yields exactly the same results as QEP-bL ($\alpha=0$) and QEP-sL.

Finally, we remark that the precessional motion in Cartesian coordinates takes the form
\begin{equation}
  \mathbf{r}_1 = R \begin{pmatrix} \sin(\theta) \cos(\dot{\phi}_{\pm} t)\\ \sin(\theta) \sin(\dot{\phi}_{\pm} t) \\ \cos(\theta) \end{pmatrix},
\end{equation}
where $\theta$ and $\dot{\phi}_{\pm}$ are constant. Because $2\cos(\dot{\phi}_{\pm} t) = \mathrm e^{\mathrm i\dot{\phi}_{\pm} t} + \mathrm e^{-\mathrm i\dot{\phi}_{\pm} t}$ and $2\mathrm i\sin(\dot{\phi}_{\pm} t) = \mathrm e^{\mathrm i\dot{\phi}_{\pm} t} - \mathrm e^{- \mathrm i\dot{\phi}_{\pm} t}$, the Fourier spectrum of this motion has the frequencies $\pm \dot{\phi}_{\pm}$. When the motion is analyzed in Fourier space, we thus obtain a total of four frequencies for the two possible precessional motions. 
Among these frequencies we may, for example, calculate
\begin{equation}
  (-\dot{\phi}_{-}) - \dot{\phi}_{+}  = |\dot{\phi}_{-}| - |\dot{\phi}_{+}| =  \omega_{\text{sc}},
\end{equation}
which corresponds to the difference in excitation energies $|\dot{\phi}_{-}|$ and $|\dot{\phi}_{+}|$. This quantity has the virtue of being insensitive to the potential energy surface $V(\theta)$ as long as it is cylindrically symmetric.

\clearpage

\begin{table}
\centering
\caption{NEO and QEP precessional ($\exclaba$ and $\exclabb$) and stretching ($\exclabc$) frequencies in units of cm$^{-1}$ for the hydrogen nucleus in  HCN  as a function of magnetic field strength. All calculations at the Hartree--Fock level of theory as described in the text.}
\begin{tabular}{cc|cccc}
$B/B_0$ & Mode & NEO-TDHF & QEP-sL &  QEP-bL & QEP-nL \\ \hline
\multirow{3}{*}{0.0} & $\exclaba$ & 972 & 753 & 753 & 753 \\
 & $\exclabb$ & 972 & 753 & 753 & 753 \\ 
 & $\exclabc$ & 3297 & 3386 & 3386 & 3386 \\  \hline
\multirow{3}{*}{0.1} & $\exclaba$ & 979 &  759 & 755 & 761 \\
 & $\exclabb$ & 985 & 764 &  767 & 761 \\ 
 & $\exclabc$ & 3312 & 3399 & 3399 & 3399 \\  \hline
 \multirow{3}{*}{0.2} & $\exclaba$ & 1006 & 780 & 773 & 785 \\
 & $\exclabb$ & 1017 &  790 & 797 & 785 \\ 
 & $\exclabc$ & 3354 & 3436 & 3436 & 3436 \\  \hline
 \multirow{3}{*}{0.3} & $\exclaba$ & 1046 & 813 & 803 & 821 \\
 & $\exclabb$ & 1063 & 828 & 839 & 821 \\ 
 & $\exclabc$ & 3419 & 3496 & 3496 & 3496 \\  \hline
 \multirow{3}{*}{0.4} & $\exclaba$ & 897 & 604 &  604 & 627 \\
 & $\exclabb$ & 945 & 651 & 652 & 627 \\ 
 & $\exclabc$ & 3543 & 3617 & 3617 & 3617 \\  \hline
 \multirow{3}{*}{0.5} & $\exclaba$ & 1082 & 840 & 837 & 867 \\
 & $\exclabb$ & 1137 & 894 & 897 & 867 \\ 
 & $\exclabc$ & 3650 & 3717 & 3717 & 3717 \\  \hline
 \multirow{3}{*}{0.6} & $\exclaba$ & 1257 & 1038 & 1032 & 1068 \\
 & $\exclabb$ & 1319 & 1098 & 1104 & 1068 \\ 
 & $\exclabc$ & 3763 & 3823 & 3823 & 3823 \\  \hline
 \multirow{3}{*}{0.7} & $\exclaba$ & 1426 & 1217 & 1209 & 1250 \\
 & $\exclabb$ & 1495 & 1285 & 1293 & 1250 \\ 
 & $\exclabc$ & 3878 & 3930 & 3930 & 3930 \\  \hline
 \multirow{3}{*}{0.8} & $\exclaba$ & 1591 & 1387 & 1377 & 1424 \\
 & $\exclabb$ & 1668 & 1462 & 1472 & 1424 \\ 
 & $\exclabc$ & 3995 & 4038 & 4038 & 4038 \\  \hline
 \multirow{3}{*}{0.9} & $\exclaba$ & 1752 & 1552 & 1540 & 1593 \\
 & $\exclabb$ & 1837 & 1634 & 1647 & 1593 \\ 
 & $\exclabc$ & 4111 & 4144 & 4144 & 4144 \\  \hline
 \multirow{3}{*}{1.0} & $\exclaba$ & 1911 & 1716 & 1701 & 1760 \\
 & $\exclabb$ & 2003 & 1805 & 1820 & 1760 \\ 
 & $\exclabc$ & 4226 & 4248 & 4248 & 4248 \\ \hline
\end{tabular}
\label{table_hcn}
\end{table}
\begin{table}
\centering
\caption{NEO and QEP precessional ($\exclaba$ and $\exclabb$) and stretching ($\exclabc$) frequencies 
in units of cm$^{-1}$
for the hydrogen nucleus in FHF$^-$ as a function of magnetic field strength.  All calculations at the Hartree--Fock level of theory as described in the text.}
\begin{tabular}{cc|cccc}
$B/B_0$ & Mode & NEO-TDHF & QEP-sL &  QEP-bL & QEP-nL \\ \hline
\multirow{3}{*}{0.0} & $\exclaba$ & 1438 & 1476 & 1476 & 1476 \\
 & $\exclabb$ & 1438 & 1476 & 1476 & 1476 \\ 
 & $\exclabc$ & 1754 & 900 & 900 & 900 \\  \hline
\multirow{3}{*}{0.1} & $\exclaba$ & 1442 & 1481 & 1480 & 1486 \\
 & $\exclabb$ & 1452 & 1491 & 1492 & 1486 \\ 
 & $\exclabc$ & 1758 & 905 & 905 & 905 \\  \hline
\multirow{3}{*}{0.2} & $\exclaba$ & 1465 & 1506 & 1504 & 1516 \\
 & $\exclabb$ & 1485 & 1526 & 1528 & 1516 \\ 
 & $\exclabc$ & 1771 & 919 & 919 & 919 \\  \hline
\multirow{3}{*}{0.3} & $\exclaba$ & 1503 & 1548 & 1545 & 1563 \\
 & $\exclabb$ & 1534 & 1577 & 1581 & 1563 \\ 
 & $\exclabc$ & 1791 & 941 & 941 & 941 \\  \hline
 \multirow{3}{*}{0.4} & $\exclaba$ & 1557 & 1604 & 1599 & 1623 \\
 & $\exclabb$ & 1597 & 1643 & 1647 & 1623 \\ 
 & $\exclabc$ & 1818 & 968 & 968 & 968 \\  \hline
\multirow{3}{*}{0.5} & $\exclaba$ & 1622 & 1669 & 1664 & 1693 \\
 & $\exclabb$ & 1671 & 1717 & 1723 & 1693 \\ 
 & $\exclabc$ & 1850 & 999 & 999 & 999 \\  \hline
 \multirow{3}{*}{0.6} & $\exclaba$ & 1696 & 1739 & 1732 & 1768 \\
 & $\exclabb$ & 1755 & 1796 & 1804 & 1768 \\ 
 & $\exclabc$ & 1884 & 1031 & 1031 & 1031 \\  \hline
\multirow{3}{*}{0.7} & $\exclaba$ & 1777 & 1810 & 1801 & 1843 \\
 & $\exclabb$ & 1845 & 1876 & 1885 & 1843 \\ 
 & $\exclabc$ & 1921 & 1063 & 1063 & 1063 \\  \hline
 \multirow{3}{*}{0.8} & $\exclaba$ & 1862 & 1880 & 1869 & 1916 \\
 & $\exclabb$ & 1939 & 1953 & 1965 & 1916 \\ 
 & $\exclabc$ & 1959 & 1094 & 1094 & 1094 \\  \hline
\multirow{3}{*}{0.9} & $\exclaba$ & 1949 & 1946 & 1933 & 1986 \\
 & $\exclabb$ & 2034 & 2027 & 2041 & 1986 \\ 
 & $\exclabc$ & 1996 & 1122 & 1122 & 1122 \\  \hline
\multirow{3}{*}{1.0} & $\exclaba$ & 2036 & 2009 & 1993 & 2051 \\
 & $\exclabb$ & 2127 & 2095 & 2112 & 2051 \\ 
 & $\exclabc$ & 2031 & 1148 & 1148 & 1148 \\  \hline 
\end{tabular}
\label{table_fhf}
\end{table}


\end{document}


\title{Time-Dependent Nuclear-Electronic Orbital Hartree--Fock Theory in a Strong Uniform Magnetic Field\\-- Supporting Information}

\author{Tanner Culpitt}
\email{t.p.culpitt@kjemi.uio.no}
\affiliation
{Hylleraas Centre for Quantum Molecular Sciences,  Department of Chemistry, 
University of Oslo, P.O. Box 1033 Blindern, N-0315 Oslo, Norway}
\author{Laurens D. M. Peters}
\affiliation
{Hylleraas Centre for Quantum Molecular Sciences,  Department of Chemistry, 
University of Oslo, P.O. Box 1033 Blindern, N-0315 Oslo, Norway}
\author{Erik I. Tellgren}
\affiliation
{Hylleraas Centre for Quantum Molecular Sciences,  Department of Chemistry, 
University of Oslo, P.O. Box 1033 Blindern, N-0315 Oslo, Norway}
\author{Trygve Helgaker}
\affiliation
{Hylleraas Centre for Quantum Molecular Sciences,  Department of Chemistry, 
University of Oslo, P.O. Box 1033 Blindern, N-0315 Oslo, Norway}

\maketitle

\clearpage

\begin{table}
\centering
\caption{Geometries of the HCN molecule at a given magnetic field strength. The molecule is oriented along the $z$-axis, and coordinates are given in Bohr.}
\begin{tabular}{cc|c}
$B/B_0$ & Atom & $z$-coordinate \\ \hline
\multirow{3}{*}{0.0} & $\text{C}$ & -1.02659 \\
 & $\text{N}$ & 1.09768 \\ 
 & $\text{H}$ & -3.02413 \\  \hline
\multirow{3}{*}{0.1} & $\text{C}$ & -1.025263  \\
 & $\text{N}$ & 1.096217 \\ 
 & $\text{H}$ & -3.019921 \\  \hline
 \multirow{3}{*}{0.2} & $\text{C}$ & -1.025298 \\
 & $\text{N}$ & 1.088201 \\ 
 & $\text{H}$ & -3.011652 \\  \hline
 \multirow{3}{*}{0.3} & $\text{C}$ & -1.025491 \\
 & $\text{N}$ & 1.075810 \\ 
 & $\text{H}$ & -2.999070 \\  \hline
 \multirow{3}{*}{0.4} & $\text{C}$ & -1.279215 \\
 & $\text{N}$ & 1.557852 \\ 
 & $\text{H}$ & -3.227468 \\  \hline
 \multirow{3}{*}{0.5} & $\text{C}$ & -1.276997 \\
 & $\text{N}$ &  1.532329 \\ 
 & $\text{H}$ & -3.204061 \\  \hline
 \multirow{3}{*}{0.6} & $\text{C}$ & -1.274487 \\
 & $\text{N}$ & 1.505314 \\ 
 & $\text{H}$ & -3.179585 \\  \hline
 \multirow{3}{*}{0.7} & $\text{C}$ & -1.271681 \\
 & $\text{N}$ & 1.477754 \\ 
 & $\text{H}$ & -3.154774 \\  \hline
 \multirow{3}{*}{0.8} & $\text{C}$ & -1.268725 \\
 & $\text{N}$ & 1.450144 \\ 
 & $\text{H}$ & -3.130201 \\  \hline
 \multirow{3}{*}{0.9} & $\text{C}$ & -1.265617 \\
 & $\text{N}$ & 1.422948 \\ 
 & $\text{H}$ & -3.106095 \\  \hline
 \multirow{3}{*}{1.0} & $\text{C}$ & -1.262466 \\
 & $\text{N}$ & 1.396411 \\ 
 & $\text{H}$ & -3.082701 \\ \hline
\end{tabular}
\label{table_hcn}
\end{table}
\begin{table}
\centering
\caption{Geometries of the FHF$^-$ molecule at a given magnetic field strength. The molecule is oriented along the $z$-axis, and coordinates are given in Bohr.}
\begin{tabular}{cc|c}
$B/B_0$ & Atom & $z$-coordinate \\ \hline
\multirow{3}{*}{0.0} & $\text{F}$ & -2.118990 \\
 & $\text{F}$ & 2.118990 \\ 
 & $\text{H}$ & 0.000000 \\  \hline
\multirow{3}{*}{0.1} & $\text{F}$ & -2.117415  \\
 & $\text{F}$ & 2.117415 \\ 
 & $\text{H}$ & 0.000000 \\  \hline
 \multirow{3}{*}{0.2} & $\text{F}$ & -2.112764 \\
 & $\text{F}$ & 2.112764 \\ 
 & $\text{H}$ & 0.000000 \\  \hline
 \multirow{3}{*}{0.3} & $\text{F}$ & -2.105268 \\
 & $\text{F}$ & 2.105268 \\ 
 & $\text{H}$ & 0.000000 \\  \hline
 \multirow{3}{*}{0.4} & $\text{F}$ & -2.095284 \\
 & $\text{F}$ & 2.095284 \\ 
 & $\text{H}$ & 0.000000 \\  \hline
 \multirow{3}{*}{0.5} & $\text{F}$ & -2.083252 \\
 & $\text{F}$ & 2.083252 \\ 
 & $\text{H}$ & 0.000000 \\  \hline
 \multirow{3}{*}{0.6} & $\text{F}$ & -2.069644 \\
 & $\text{F}$ & 2.069644 \\ 
 & $\text{H}$ & 0.000000 \\  \hline
 \multirow{3}{*}{0.7} & $\text{F}$ & -2.054906 \\
 & $\text{F}$ & 2.054906 \\ 
 & $\text{H}$ & 0.000000 \\  \hline
 \multirow{3}{*}{0.8} & $\text{F}$ & -2.039446 \\
 & $\text{F}$ & 2.039446 \\ 
 & $\text{H}$ & 0.000000 \\  \hline
 \multirow{3}{*}{0.9} & $\text{F}$ & -2.023599 \\
 & $\text{F}$ & 2.023599 \\ 
 & $\text{H}$ & 0.000000 \\  \hline
 \multirow{3}{*}{1.0} & $\text{F}$ & -2.007638 \\
 & $\text{F}$ & 2.007638 \\ 
 & $\text{H}$ & 0.000000 \\ \hline
\end{tabular}
\label{table_fhf}
\end{table}